\documentstyle[fullpage,epsf]{article}

\newtheorem{satz}{Theorem}

\newtheorem{lemma}[satz]{Lemma}
\newtheorem{definition}[satz]{Definition}
\newtheorem{prop}[satz]{Proposition}
\newcommand{\nat}{{\rm I \! N}}
\newcommand{\old}[1]{}
\newcommand{\rnn}{{{\rm I \! R}^{+}_{0}}}
\newcommand{\qed}{\hfill$\Box$}
\newcommand{\clique}[1]{\mbox{$E_{#1}$}}
\newcommand{\parti}{\stackrel{.}{\bigcup}}

\begin{document}

\title{An Exact Algorithm for Higher-Dimensional Orthogonal Packing\thanks{
A previous extended abstract version of this paper appears in
{\em Algorithms -- ESA'97}~\cite{esa}}}

\author{S\'andor P. Fekete\\
Department of Mathematical Optimization\\
Braunschweig University of Technology\\
D--38106 Braunschweig\\
GERMANY\\
        {\tt s.fekete@tu-bs.de}\and
        J\"org Schepers\thanks{
Supported by the German Federal Ministry of Education,
Science, Research and Technology (BMBF, F\"orderkennzeichen 01~IR~411~C7).}\\
        IBM Germany\\
        Gustav-Heinemann-Ufer 120/122\\
        D--50968 K\"oln\\
        GERMANY\\
        {\tt schepers@de.ibm.com}\and
Jan C. van der Veen\\
Department of Mathematical Optimization\\
Braunschweig University of Technology\\
D--38106 Braunschweig\\
GERMANY\\
        {\tt j.van-der-veen@tu-bs.de}
}
 
\date{}
\maketitle 

\begin{abstract}

Higher-dimensional orthogonal packing problems have a wide range of 
practical applications, including packing, cutting, and scheduling.
Combining the use of our data structure for characterizing
feasible packings with our new classes of lower bounds, and 
other heuristics, we develop a two-level tree search algorithm
for solving higher-dimensional packing problems to optimality. 
Computational results are reported, including optimal
solutions for all two--dimensional test problems from recent literature.

This is the third in a series of articles describing
new approaches to higher-dimensional packing.
\end{abstract}

\section{Introduction}
Most combinatorial optimization problems that need to be solved in
practical applications turn out to be members of
the class of NP-hard problems. Algorithmic research of several decades
has provided strong evidence that for all
of these problems, it is highly unlikely that there is a {\em polynomial
algorithm}: Such an algorithm is guaranteed to find an optimal solution
in time that even in the worst case can be bounded by a polynomial
in the size of the input. If no such bound can be guaranteed,
the necessary time for solving instances tends to grow very fast 
as the instance size increases. That is why NP-hard problems have
also been dubbed ``intractable''. See the classical monograph~\cite{GAJO79}
for an overview.

When confronted with an NP-hard problem, there are several
ways to deal with its computational difficulty:

{\em We can look for a different problem.} 

While this way out may
be quite reasonable in a theoretical context, it tends to work less well
when a problem arises in practical applications that have to be
solved somehow.

{\em We can look for special properties of a problem instance
or relax unimportant constraints in order to get a polynomial algorithm.} 

Unfortunately, practical instances and their additional constraints
tend to be more difficult at a second glance, rather than simpler.

{\em We can look for a good solution instead of an optimal one.}

This approach has received an increasing amount of attention
over recent years.  In particular, there has been a tremendous amount
of research dealing with polynomial time approximation algorithms
that are guaranteed to find 
a solution within a fixed multiplicative
constant of the optimum. See the book~\cite{Hoch} for an overview.

{\em We can look for an optimal solution without a 
bound on the runtime.}

While the time for finding an optimal solution may be quite long
in the worst case, a good understanding of the underlying mathematical
structure may allow it to find an optimal solution (and prove it)
in reasonable time for a large number of instances. A good example
of this type can be found in \cite{GRO80}, where the exact solution
of a 120-city instance of the Traveling Salesman Problem
is described. In the meantime, benchmark instances
of size up to 13509 and 15112 cities have been solved to 
optimality \cite{13509},
showing that the right mathematical tools and sufficient computing power
may combine to explore search spaces of tremendous size. In this sense,
``intractable'' problems may turn out to be quite tractable.


In this paper, we consider a class of problems that is not only
NP-hard, but also difficult in several other ways.
Packing rectangles into a container arises in many industries,
where steel, glass, wood, or textile materials are cut, but it also occurs
in less obvious contexts, such as
machine scheduling or optimizing the layout of advertisements in newspapers.
The three-dimensional problem is important for practical
applications as container
loading or scheduling with partitionable resources.
Other applications arise from allocating jobs to reconfigurable
computer chips---see~\cite{Teich}.
For many of these problems, objects must be positioned with
a fixed orientation; this is a requirement that we will
assume throughout the paper.
The {\em $d$-dimensional orthogonal knapsack problem} (OKP-$d$)
requires to select a most valuable subset $S$ from a given set of
boxes, such that $S$ can be packed into the container.
Being a generalization of the one-dimensional bin packing problem,
the OKP-$d$ is NP-complete in the strict sense.
Other NP-hard types of packing problems include the 
{\em strip packing problem} (SPP), where we need to minimize
the height of a container of given width, such that a given set of boxes
can be packed, and the {\em orthogonal bin packing problem} (OBPP)
where we have a supply of containers of a given size and need
to minimize the number of containers that are needed for
packing a set of boxes. The decision version of these problems
is called the {\em Orthogonal Packing Problem} (OPP), where
we have to decide whether a given set of boxes fits into
a container. 

Relatively few authors have dealt with the exact 
solution of orthogonal knapsack problems.
All of them focus on the problem in two dimensions.
One of the reasons is the difficulty of giving a
simple mathematical description of the set of feasible
packings: As soon as one box is packed into the container,
the remaining feasible space is no longer convex, 
excluding the straightforward application of integer
programming methods.
Bir\'{o} and Boros (1984) \cite{BIBO84} give a 
characterization of non--guillotine patterns using
network flows but derived no algorithm. Dowsland (1987) \cite{DOWS87} 
proposes an exact algorithm
for the case that all boxes have equal size. 
Arenales and Morabito (1995) \cite{ARMO95} extend an approach
for the guillotine problem to cover a certain type of non--guillotine patterns. 
So far, only three exact algorithms have been proposed 
and tested for the general case. Beasley (1985) \cite{BEA85}
and Hadjiconstantinou and Christofides (1995) \cite{HACHR95} 
give different 0--1 integer
programming formulations of this problem. 
Even for small problem instances, 
they have to consider very large 0--1 programs,
because the number of variables depends on the size of the container that is to be
packed. The largest instance that is solved in either article has 9 out of 22 boxes packed into
a 30 $\times$ 30 container. After an initial reduction phase,
Beasley gets a 0-1 program with more than 8000 variables and more than
800 constraints; the program by   
Hadjiconstantinou and Christofides still contains more than 
1400 0--1 variables and over 5000 constraints. 
>From Lagrangean relaxations, they derive upper bounds for a 
branch--and--bound algorithm, which are
improved using subgradient optimization.
The process of traversing the search tree corresponds to the iterative 
generation of an optimal packing. 
More recent work by Caprara and Monaci (2004)~\cite{CM04} on
the two-dimensional knapsack problem
uses our previous results (cited as \cite{esa,pack3}
as the most relevant reference for comparison; we compare our results 
and approaches later
in this paper and discuss how a combination of our methods
may lead to even better results.

Other research on the related problem
of two- and three-dimensional bin packing has been presented:
Martello and Vigo (1996) \cite{MAVI96}
consider the two-dimensional case,
while
Martello, Pisinger, and Vigo (1997) \cite{MAPV97} deal
with three-dimensional bin packing. We discuss aspects
of those papers in 
\cite{pack2b}, when considering
bounds for higher-dimensional packing problems.
Padberg~(2000) \cite{Pad00} gives a mixed integer
programming formulation for three-dimensional packing problems,
similar to the one anticipated by the second author
in his thesis \cite{SEP97}.
Padberg expresses the hope that using a number of techniques from
branch-and-cut will be useful; however, he does not provide
any practical results to support this hope.

In our papers \cite{pack1b,pack2b}, 
we describe a different approach to characterizing
feasible packings and constructing optimal solutions.
We use a graph-theoretic characterization of the relative position
of the boxes in a feasible packing \cite{pack1b}. 
Combined with good heuristics for dismissing infeasible subsets
of boxes that are described in \cite{pack2b}, this characterization
can be used to develop a two-level tree search. In this
third paper of the series, we describe how
this exact algorithm can be implemented.
Our computational results show that our code outperforms 
previous methods by a wide margin.
It should be noted that our approach has been used and extended
in the practical context of reconfigurable computing \cite{Teich},
which can be interpreted as packing in three-dimensional space,
with two coordinates describing chip area and one coordinate
describing time. Order constraints for the temporal order
are of vital importance in this context; as it turns out, our
characterization of feasible packings is particularly suited
for taking these into account. See our followup paper
\cite{pack4} for a description of how to deal with higher-dimensional
packing with order constraints.

The rest of this paper is organized as follows: after recalling some
basics from our papers \cite{pack1b,pack2b,pack1,pack2} in Section~2,
we give detailed account of our approach for handling OPP instances
in Section~3. This analysis includes a description
of how to apply graph theoretic characterizations of
interval graphs to searching for optimal packings.
Section~4 provides details of our branch-and-bound framework
and the most important subroutines. In Section~5, we discuss
our computational results. Section~6 gives a brief description
of how our approach can be applied to other types of packing problems.

\section{Preliminaries}
We are given a finite set $V$ of $d$-dimensional rectangular boxes
with ``sizes''  $w(v) \in \rnn^{d}$ and ``values'' 
$c(v) \in \rnn$ for $v \in V$. As we are considering fixed orientations,
$w_i(v)$ describes the size of box $v$ in the $x_i$-direction.
The objective of the {\em $d$-dimensional orthogonal knapsack problem (OKP-$d$)}
 is
to maximize the total value of a subset $V' \subseteq V$ fitting into the
container $C$ and to find a complying packing.
Closely related is the
{\em $d$-dimensional orthogonal packing problem (OPP-$d$)}, which is
to decide whether a given set of boxes $B$ fits into a unit size container, and
to find a complying packing whenever possible.

For a $d$-dimensional packing, we consider the projections of the boxes
onto the $d$ coordinate axes $x_i$. Each of these projections induces
a graph $G_i$: two boxes are adjacent in $G_i$, if and only if
their $x_i$ projections overlap. A set of boxes
$S \subseteq V$ is called
{\em $x_i$--feasible}, if the boxes in $S$
can be lined up along the $x_i$--axis
without exceeding the $x_i$-width of the container.

As we show in \cite{pack1b,pack1}, we have the following characterization
of feasible packings:

\begin{satz}
A set of boxes $V$ can be packed into a container, if and only if there
is a set of $d$ graphs $G_i=(V, E_i), i=1,\ldots,d$
with the following properties:
\begin{eqnarray*}
P1:&& \mbox{the graphs $G_{i}:=(V,E_{i})$ are interval graphs.}
\\
P2:&& \mbox{each stable set } S \mbox{ of } G_{i} \mbox{ is $x_i$--feasible}.
\\
P3:&& \bigcap_{i=1}^{d} E_{i} = \emptyset.
\end{eqnarray*}
\end{satz}

A set $E=(E_1,\ldots,E_d)$ of edges is called a {\em packing class}
for $(V,w)$,
if and only if it satisfies the conditions $P1$, $P2$, $P3$.
Note that when constructing a packing from a packing class, some
edges may be added in case of a degenerate packing; see
Ferreira and Oliveira \cite{FO} for such an example.
This does not impede the correctness of the theorem or its applicability.

\section{Solving Orthogonal Packing Problems}\label{oppsec}
For showing feasibility of any solution to a packing problem,
we have to prove that a particular set of boxes fits into
the container. This subproblem is called the
{\em orthogonal packing problem (OPP)}.

In order to get a fast positive answer, we can try to find
a packing by means of a heuristic. A fast way to get a negative
answer has been described in our paper~\cite{pack2b}:
Using a selection of bounds (conservative scales), we can try to apply 
the volume criterion to show that there cannot be a feasible packing.

In this section, we discuss the case that both of these
easy approaches fail. Because the OPP is NP-hard in the strong sense,
it is reasonable to use enumerative methods. As we showed
in our paper~\cite{pack1b}, the existence of a packing is equivalent
to the existence of a packing class. Furthermore, we have shown
that a feasible packing can be constructed from a packing class in 
time that is linear in the number of edges.
This allows us to search for a packing class, instead of a packing.
As we will see in the following, the advantage of this approach
lies not only in exploiting the symmetries discussed in \cite{pack1b,pack1},
but also in the fact that the structural properties of packing classes
give rise to very efficient rules for identifying irrelevant
portions of the search tree.

\subsection{Basic Idea of the Enumeration Scheme}
The enumeration of packings described by Beasley
in \cite{BEA85} emulates the intuitive idea of packing objects
into a box: Each branching corresponds to
placing a box at a particular position in the container, 
or disallowing this placement.
Thus, each search node corresponds to a partial packing that
is to be augmented to a complete packing.
Our enumeration of packing classes is more abstract than that:
At each branching we decide whether two boxes $b$ and $c$
overlap in their projection onto the $i$-axis, so that the 
edge $e := bc$ is contained in the $i$th component graph 
of the desired packing class $E$.
Accordingly, in the first resulting subtree, we only search for
packing classes $E$ with $e \in E_{i}$; in the second, we only search
for $E$ with $e \notin E_{i}$.
Hence, the resulting ``incomplete packing classes''
do {\em not} correspond to packing classes of subsets of boxes,
instead they are (almost) arbitrary tuples of edges.

More precisely, we will store the ``necessary''
and ``excluded '' edges for each node $N$ of the search tree and
each coordinate direction $i$ in two data structures
${\cal E}^{N}_{+,i}$ and ${\cal E}^{N}_{-,i}$.
Therefore, the  search space for $N$ contains precisely
the packing classes that satisfy the condition
\begin{equation} \label{schachtel}
{\cal E}_{+,i}^{N} \subseteq E_{i} \subseteq \overline{{\cal E}_{-,i}^{N}}, \, i \in \{1, \dots, d\}, 
\end{equation}
where $\overline{{\cal E}_{-,i}^{N}}$ is the complement of ${{\cal E}_{-,i}^{N}}$. Summarizing, we write
$$
{\cal E}^{N}_{+} :=
({\cal E}^{N}_{+,1}, \dots, 
{\cal E}^{N}_{+,d} ),
\quad
{\cal E}^{N}_{-} :=
({\cal E}^{N}_{-,1}, \dots, 
{\cal E}^{N}_{-,d} ),
\quad 
{\cal E}^{N} :=
({\cal E}^{N}_{+},
{\cal E}^{N}_{-})
$$
and denote by ${\cal L}({\cal E}^{N})$
the search space for $N$; by virtue of (\ref{schachtel}), this search space
is only determined by ${\cal E}^{N}$.
${\cal E}^{N}$ is called the {\em search information} of node $N$, because 
this tuple of data structures represents the information that
is currently known about the desired packing class.

An important part of the procedure consists in using the characteristic
properties $P1$, $P2$, $P3$ for increasing the information
on the desired packing class that is contained in ${\cal E}^{N}$.
For example, let the edge
$e$ be contained in ${\cal E}_{+,i}$ for all $i \ne k$.
Furthermore, let 
$E \in {\cal L}({\cal E}^{N})$.  
For $i \ne k$, we have $e \in E_{i}$ because of 
${\cal E}^{N}_{+,i} \subseteq E_{i}$.
Because of $P3$, the intersection of all $E_{i}$ must be empty,
implying $e \notin E_{k}$.
Therefore, ${\cal E}^{N}_{-,k}$ can be augmented by $e$ 
without changing the search space. 
Similar augmentation rules can be described for 
$P1$ and $P2$.

Depending on whether the edge $e$ is added to ${\cal E}^{N}_{+,i}$
or to ${\cal E}^{N}_{-,i}$, we describe augmentations
of ${\cal E}^{N}$ by the triples $(e,+,i)$ or $(e,-,i)$.

Because it suffices to find a single packing class,
these augmentations may reduce the search space, as long as it is guaranteed
that not all packing classes are removed from it.
This fact allows us to exploit certain symmetries.
Thus, we use {\em feasible augmentations} of
${\cal E}^{N}$ in the sense that a nonempty search space 
${\cal L}({\cal E}^{N})$ 
stays nonempty after the augmentation.

When augmenting 
${\cal E}^{N}$ we follow two objectives:
\begin{enumerate}
\item
obtain a packing class in ${\cal E}^{N}_{+}$, 
\item 
prove that every augmentation of 
${\cal E}^{N}_{+}$ to a packing class has to 
use ``excluded '' edges from 
${\cal E}^{N}_{-}$.
\end{enumerate}

In the first case, our tree search has been successful.
In the second case, the search on the current subtree may be terminated,
because the search space is empty.
Otherwise, we have to continue branching
until one of the two objectives is reached.

\subsection{Excluded Induced Subgraphs} \label{induzsubsection}
For our algorithm, we need three components:
a test 
``Is ${\cal E}^{N}_{+}$ a packing class?'', a
sufficient criterion that 
${\cal E}^{N}_{+}$ has no feasible augmentation,
and a construction method for feasible augmentations.
As we describe in our paper~\cite{pack1b,pack1},
all three of these components can be reduced to 
identifying or avoiding particular induced subgraphs
in the portions of $E$ that are fixed by ${\cal E}^{N}$.

As we have already seen, it is easy to determine all
edges that are excluded by condition $P3$.
By performing these augmentations of
${\cal E}^{N}_{-}$ immediately, we can guarantee that 
$P3$ is satisfied. Thus we will assume in the following that $P3$ is satisfied.
Furthermore we will implicitly refer to the current
search node $N$ and abbreviate 
${\cal E}^{N}$ by ${\cal E}$.

$P2$ explicitly excludes certain induced subgraphs:
$i$-infeasible stable sets, i.\,e., $i$-infeasible cliques in the complement
of each component graph.

In order to formulate $P1$ in terms of excluded induced
subgraphs, we recall the following Theorems~\ref{ivchar1} and \ref{kompchar}
-- see the book by Golumbic\cite{GOL80},
as well as a resulting linear-time algorithm by
Korte and M\"ohring \cite{KOMO89}. The following terminology is used:

\begin{definition}
For a graph $G:=(V,E)$, a set $F \subseteq V^{2}$ of directed edges
is an {\em orientation of $G$}, iff
$$
\forall b,c \in V: \, bc \in E \, \Longleftrightarrow
(\vec{bc} \in F \wedge \vec{cb} \notin F)
\vee
(\vec{cb} \in F \wedge \vec{bc} \notin F)
$$
holds.
An orientation $F$ of a graph $(V,E)$ is called {\em transitive},
if in addition, $$
\forall b,c,z \in V: \, \vec{bc} \in F \, \land \, \vec{cz} \in F \Rightarrow \vec{bz} \in F
$$
holds.

A graph is called a {\em comparability graph}, iff it has a transitive
orientation. 

For a cycle $C := [b_{0}, \dots, b_{k-1}, b_{k}=b_{0}]$ of length $k$,
the edges $b_{i}b_{j}, i,j \in \{0,\dots,k-1\}$ 
with $(|i-j| \mbox{ mod } k) > 1$ are called {\em chords}; the chords
$b_{i}b_{j}, i,j \in \{0,\dots,k-1\}$ with $(|i-j| \mbox{ mod } k) = 2$
are called {\em $2$-chords} of $C$.
A cycle is ($2$-) {\em chordless}, iff it does not have any
($2$-) chords.

A graph $G=(V,E)$ is a {\em cocomparability graph}, if the complement
graph $G=(V,\overline{E})$ is a comparability graph.
\end{definition}

\begin{satz}[Gilmore and Hoffman 1964] \label{ivchar1}
A cocomparability graph is an interval graph, iff
it does not contain the chordless cycle $C_{4}$ of length 4 
as an induced subgraph.
\end{satz}

\begin{satz}[Ghouil\`{a}-Houri 1962, Gilmore and Hoffman 1964] \label{kompchar}
A graph is a comparability graph, iff it does not contain a $2$-chordless 
cycle of odd length.
\end{satz}


Thus, ${\cal E}_{+}$ is a packing class, if for all 
$i \in \{1,\dots,d\}$ the following holds
(recall that $P3$ is assumed to be satisfied):              
\begin{enumerate}
\item $(V,{\cal E}_{+, i})$ does not contain a $C_{4}$ as an induced
subgraph.
\item $(V,\overline{{\cal E}_{+, i}})$ does not contain
an odd $2$-chordless cycle.
\item $(V,\overline{{\cal E}_{+, i}})$ does not contain
an $i$-infeasible clique.
\end{enumerate}

With the help of this characterization, we get a stop criterion
for subtrees.
Because only those edges can be added to 
${\cal E}_{+, i}$ that are not
in ${\cal E}_{-, i}$,
${\cal E}_{+}$ cannot be augmented to a packing class, if
for $i \in \{1,\dots,d\}$ one of the following conditions
holds:
\begin{enumerate}
\item $(V,{\cal E}_{+, i})$ contains a $C_{4}$ as an induced subgraph,
with both chords lying in ${\cal E}_{-, i}$.
\item $(V,{\cal E}_{-, i})$ contains an odd $2$-chordless cycle, with all
its $2$-chords lying in ${\cal E}_{+, i}$.
\item $(V,{\cal E}_{-, i})$ contains a $i$-infeasible clique.
\end{enumerate}
Suppose that except for one edge $e$,
one of these excluded configurations is contained in
${\cal E}$.
Because of condition~\ref{schachtel}., the corresponding
incomplete induced subgraph is contained in the $i$th
component graph of each packing class
$E \in {\cal L}({\cal E})$.
Because completing the excluded subgraph would contradict condition
$P1$ or $P2$, the membership of $e$ in 
$E_{i}$ is determined.
The resulting forced edges can be added to
${\cal E}_{+,i}$ or
to ${\cal E}_{-,i}$ without decreasing the search space.

{\bf Example:}
If $(V,{\cal E}_{+, i})$ contains an induced $C_{4}$, for which one
chord is contained in
${\cal E}_{-, i}$, for any packing class of the search space,
the other chord $e$ must be contained in the $i$th component graph.
Thus the augmentation $(e,+,i)$ is feasible, but not 
the augmentation $(e,-,i)$. 

In this way, we can reduce the search for a feasible augmentation
to the search for incomplete excluded configurations.
In the next section, we will relax the condition that only one 
edge is missing from a configuration, and only require that the
missing edges are equivalent in a particular sense.

\subsection{Isomorphic Packing Classes} \label{isosection}
When exchanging the position of two boxes with identical sizes
in a feasible packing, we obtain another feasible packing.
Similarly, we can permute equal boxes in a packing class:
\begin{satz} \label{permppc}
Let $E$ be a packing class for $(V,w)$ and 
$\pi:V \rightarrow V$ be a permutation with
\begin{equation}
\forall b \in V: \, w(b) = w(\pi(b)). \label{typtreu}
\end{equation}
Then the 
$d$-tuple of edges in $E^{\pi}$ that is given by 
$$
\forall b,c \in V \, \forall i \in \{1,\dots,d\} \quad
bc \in E_{i} \, \Leftrightarrow \, \pi(b)\pi(c) \in E^{\pi}_{i},
$$
is a packing class.
\end{satz}
{\bf Proof:} Because the structure of
the component graphs does not change, conditions $P1$ and $P3$
remain valid.
Because of (\ref{typtreu}), the weight of stable sets remains unchanged,
so that $P2$ remains valid as well.
\qed

We get a notion of isomorphism that is similar to the isomorphism
of graphs:
\begin{definition} \label{defpmciso}
Two packing classes $E$ and $E'$ are called isomorphic, 
iff there is a permutation $\pi:V \rightarrow V$
satisfying (\ref{typtreu}),
such that $E'=E^{\pi}$.
\end{definition}
Keeping only one packing class from each isomorphism class in search space
avoids unnecessary work.
For this purpose, we may assume that the ordering of equal boxes in a 
packing class follows the lexicographic order of their position vectors.
As a result, in the two-dimensional case, the leftmost and bottommost
box of a box type will have the lowest index.
This corresponds to generating packings according to
``leftmost downward placement'' in \cite{HACHR95}.
This approach cannot be used for packing classes,
because there are no longer any orientations
(left/right, up/down, etc.)

Until now, no algorithm has been found
that can decide in polynomial time whether two graphs are
isomorphic, and it has been conjectured that no such algorithm
exists (see \cite{PAPA94}, p.~291) When deciding whether
two packing classes are isomorphic, this decision has to be made repeatedly.
In addition, packing classes may only be known partially.
This makes it unlikely that there is an efficient method
for achieving optimal reduction of isomorphism.
Therefore we are content with exploiting certain cases
that occur frequently.

In the initializing phase, we may conclude by
Theorem~16 from our paper \cite{pack2b}
(corresponding to Theorem~11 in \cite{pack2}) that for a
box type $T$, there is a component graph $(V,E_{i})[T]$, for 
which there is a clique of size $k\ge 2$. 
Then we can choose the numbering of $T$, such that the first
$k$ boxes from $T$ belong to the clique.
Thus, the corresponding ${k \choose 2}$ edges can be fixed in
${\cal E}_{+,i}$.
This restriction of numbering $T$
corresponds to excluding isomorphic packing classes.

In the following, we only consider isomorphic packing classes for which
the permutation in 
Definition~\ref{defpmciso} exchanges precisely two boxes, while leaving
all other boxes unchanged. This restricted isomorphism can be checked easily.
We have to search for pairs of boxes that can be exchanged in the following 
way:
\begin{definition}
Let $(V,w)$ be an OPP instance with search information ${\cal E}$.
Two boxes $b,c \in V$ 
with $w(b)=w(c)$ are called
 {\em indistinguishable (with respect to ${\cal E}$)}, 
iff all adjacencies of $b$ and $c$ have identical search information, i.\,e.,
\begin{equation} \label{vertausch}
\forall i \in \{1,\dots,d\} \quad 
\forall \sigma \in \{+,-\} \quad 
\forall z \in V \setminus \{b,c\} : \quad
bz \in {\cal E}_{\sigma,i} \Leftrightarrow cz \in {\cal E}_{\sigma,i}. 
\end{equation}

Two edges $e,e' \in \clique{V}$ are called 
 {\em indistinguishable (with respect to ${\cal E}$)}, if
there are representations $e=bc$ and $e'=b'c'$,
such that the boxes $b$ and $b'$, as well as  $c$ and $c'$ 
are indistinguishable (with respect to ${\cal E}$). 
\end{definition}
The property of being indistinguishable is an equivalence relation
for boxes as well as for edges.

The following lemma allows it to exploit the connection between
indistinguishable edges and isomorphic packing classes:
\begin{lemma} \label{vertauschlemma}
Let $(V,w)$ be an OPP instance with search information
${\cal E}$.
Let $A$ be set of of indistinguishable edges on $V$.
Let $e$ be an arbitrary 
$e \in A$.
Then for any packing class
$E \in {\cal L}({\cal E})$ that satisfies
$A \cap E_{i} \ne \emptyset$, there is an isomorphic
packing class $E' \in {\cal L}({\cal E})$ with $e \in E'_{i}$.
\end{lemma}
{\bf Proof:}  
Let $e'$ be an edge from the set $A \cap E_{i}$.
Then $e$ and $e'$ are indistinguishable.
Hence there is a representation 
$e=bc$ and $e'=b'c'$, such that 
$b,b'$ and $c,c'$ are pairs of indistinguishable boxes.
Let $\pi$ be the permutation of $V$ that swaps $b$ and $b'$, and $c$ and $c'$,
and let $E':=E^{\pi}$.
Applying (\ref{vertausch}) twice, it follows from
$E \in {\cal L}({\cal E})$ that
$E' \in {\cal L}({\cal E})$.
\qed

\bigskip
Lemma~\ref{vertauschlemma} can be useful in two situations:
\begin{enumerate}
\item If we branch with $e \in A$ with respect to the $i$-direction,
then we may assume for all $e' \in A \,$ in the subtree
``$e \notin E_{i}$'' that $e' \notin E_{i}$:
For each packing class excluded in this way, there is an isomorphic
packing class that is contained in the search space of the 
subtree $e \in E_{i}$.
\item If during the course of our computations, we get
$A \cap {\cal E}_{+,i} \ne \emptyset$, then for an arbitrary
$e \in A$, the augmentation $(e,+,i)$ is feasible, because 
only isomorphic duplicates are lost. 
\end{enumerate}

\subsection{Pruning by Conservative Scales}
By Lemma 20 from our paper \cite{pack2b}
(Lemma 15 in \cite{pack2}), 
we can use the information given
by ${\cal E}$ to modify a given conservative scale, such that
the resulting total volume of $V$ is increased.
Before branching, we try to apply the lemma repeatedly, such that the
transformed volume exceeds the volume of the container.
In this case, the search can be stopped.

Because this reduction heuristic requires the computation of
several one-dimensional knapsack problems, it only pays off to use it 
on nodes where it may be possible to cut off large subtrees.
Therefore, we have only used it 
on nodes of depth at most 5.

\section{Detailed Description of the OPP Algorithm}

In this section, we give a detailed description of our implementation
of the OPP algorithm. We will omit the description of standard techniques
like efficient storage of sets, lists, graphs, or the implementation
of graph algorithms. The interested reader can find these
in \cite{MEHL88} and \cite{GOL80}.

\subsection{Controlling the Tree Search}

\begin{figure}[htbp] 
\fbox{\parbox{15.6cm}{
\begin{tabular}{ll}
{\bf Call: } & {\bf Solve\_OPP}($P$)\\
\\
{\bf Input: } & An OPP-$n$ instance $P:=(V,w)$. \\
\\
{\bf Output: } & 
A packing class for $(V,w)$, if there is one, and SUCCESS,
\\ & otherwise NULL. \\
\\
\end{tabular}

\begin{tabular}{rl}
1. & ${\cal N} := \{ N_{0} \}$. \\
2. & initialize ${\cal E}^{N_{0}}$. \\
3. & $(e,\sigma,i)^{N_{0}} := $ NULL.\\
4. & \\
5. & {\bf \underline{while} } ( ${\cal N} \ne \emptyset$ ) {\bf \underline{do}}\\
6. & \\
7. & $\quad$ choose $N \in {\cal N}$.\\
8. & $\quad$ ${\cal N} := {\cal N} \setminus \{N\}$.\\
9. & $\quad$ $(e,\sigma,i) := {(e,\sigma,i)}^{N}$.\\
10. & \\
11. & $\quad$ {\bf \underline{repeat}}\\
12. & $\qquad$ {\bf \underline{if}}  
({\bf Update\_$\,$searchinfo} ($P$, ${(e,\sigma,i)}$, ${\cal E}^{N}$)
$=$ EXIT ) {\bf \underline{then}} \\
13. & $\qquad \quad$ result $:=$ EXIT. \\
14. & $\qquad$ {\bf \underline{else} } \\
15. & $\qquad \quad$ result $:=$ {\bf Packingclass\_test} ($P$, ${\cal E}^{N}$, $(e,\sigma,i)$).\\
16. & $\qquad$ {\bf \underline{end if} } \\
17. & $\quad$ {\bf \underline{until} } ( result $\ne$ FIX ) \\
18. & \\
19. & $\quad$ {\bf \underline{if} } ( result = SUCCESS ) {\bf  \underline{then} \underline{return}} ${\cal E}^{N}_{+}$.\\
20. & \\
21. & $\quad$ {\bf \underline{if} } ( result = BRANCH ) {\bf  \underline{then} } \\
22. & $\qquad$ Create two new nodes $N',N''$.\\
23. & $\qquad$ ${\cal E}^{N'\,} := {\cal E}^{N}, 
 \quad \, {(e,\sigma,i)}^{N'\,} := (e,+,i)$.\\
24. & $\qquad$ ${\cal E}^{N''} := {\cal E}^{N}, 
 \quad {(e,\sigma,i)}^{N''} := (e,-,i)$.\\
25. & $\qquad$ ${\cal N} := {\cal N} \cup \{N',N''\}$.\\
26. & $\quad$ {\bf \underline{end if}} \\
27. & \\
28. & {\bf \underline{end while} }\\
29. & \\
30. & {\bf \underline{return}} NULL.\\
\end{tabular}
}}
\caption{OPP tree search}
\label{oppsearch}
\end{figure}     

The nodes of the search tree are maintained in a list ${\cal N}$.
For each node $N \in {\cal N}$, there is the search information 
${\cal E}^{N}$ (see above) and a triple
$(e,\sigma,i)^{N}$ with $e \in {V \choose 2}$, 
$\sigma \in \{+,-\}$ and $i \in \{1,\dots,d\}$.
This triple represents the new information when branching at $N$, i.\,e.,
$e \in E_{i}$ for $\sigma = +$, or 
$e \notin E_{i}$ for $\sigma = -$.

Figure~\ref{oppsearch} shows the course of the tree search.
Lines 1.\ through 3.\ initialize ${\cal N}$ with the root node
$N_{0}$. Initially, the components of
${\cal E}^{N_{0}}$ do not contain any edges.
${(e,\sigma,i)}^{N_{0}}$ is assigned a special value of ``NULL''.
 
In the while loop of lines 5.\ through 28., individual nodes are processed;
if necessary, their children are added to ${\cal N}$.
The particular branching strategy (breadth first or depth first) 
can be specified by a selection mechanism in line 7.
If line 30.\ is reached, then the whole search tree was checked
without finding a packing pattern.

Each node $N$ of the search tree is processed as follows:

In routine 
{\bf Update\_$\,$searchinfo}, the augmentation of ${\cal E}^{N}$
described by  $(e,\sigma,i)$ is carried out, as long as there
are feasible augmentations.
If it is detected that the search on $N$ can be stopped,
{\bf Update\_$\,$searchinfo} outputs the value
``EXIT''.
Otherwise, the routine terminates with ``OK''.

If {\bf Update\_$\,$searchinfo} was terminated with ``OK'',
then the routine {\bf Packingclass\_test} checks whether
${\cal E}^{N}_{+}$ already is a packing class.
In case of a positive answer, ``SUCCESS'' is output, and the algorithm
terminates in line 19.
Otherwise, there are three possibilities:
\begin{enumerate}
\item ``FIX'': The triple $(e,\sigma,i)$ was updated in
{\bf Packingclass\_test} to a new feasible augmentation
that was returned to {\bf Update\_$\,$searchinfo}.
\item ``EXIT'': ${\cal E}^{N}_{+}$ cannot be augmented
to a packing class without using edges from 
${\cal E}_{-}^{N}$. The search on this subtree is stopped.
\item ``BRANCH'': 
In lines 22.\ through 25., two children of $N$ are added to ${\cal N}$.
The triple $(e,\sigma,i)$ that was set in {\bf Packingclass\_test}
contains the branching edge $e$ and the branching direction $i$.
\end{enumerate}

\subsection{Testing for Packing Classes}
\begin{figure} 
\fbox{\parbox{15.6cm}{
\begin{tabular}{ll}
{\bf Call:} & {\bf Packingclass\_test}($P$, ${\cal E}$, $(e,\sigma,i)^{out}$)\\
\\
{\bf Input: } & An OPP-$n$ instance $P:=(V,w)$, search information ${\cal E}$. \\
\\
{\bf Output: } & $(e,\sigma,i)^{out}$, and\\
& EXIT, FIX, BRANCH, or SUCCESS \\
\\
\end{tabular}

\begin{tabular}{rl}
1. & {\bf \underline{for} } $i \in \{1,\dots,d\}$ {\bf \underline{do}}\\
2. & \\
3. & $\quad$ $A := \emptyset$.\\
4. & $\quad$ {\bf \underline{if} }  
$(V,\overline{{\cal E}_{+,i}})$ is not a comparability graph {\bf  \underline{then}}\\
5. & $\qquad$ $A$ := set of edges of the $2$-chordless odd cycle. \\
6. & $\quad$ {\bf \underline{else}} \\
7. & $\qquad$ {\bf \underline{if} }  
a maximal weighted clique in $(V,\overline{{\cal E}_{+,i}})$ is $i$-infeasible
{\bf  \underline{then}}\\
8. & $\qquad \quad$ $A$ := edge set of this clique. \\
9. & $\qquad$ {\bf \underline{else}} \\
10. & $\qquad \quad$ {\bf \underline{if} }  
$(V,{\cal E}_{+,i})$ contains an induced $C_{4}$ {\bf  \underline{then}}\\
11. & $\qquad \qquad$ $A$ := set of chords of this $C_{4}$. \\
12. & $\qquad \quad$ {\bf \underline{end if}} \\
13. & $\qquad$ {\bf \underline{end if}} \\
14. & $\quad$ {\bf \underline{end if}} \\
15. & \\
16. & \\
17. & $\quad$ {\bf \underline{if} }  
($A \ne \emptyset$) {\bf  \underline{then}}\\

18. & $\qquad$ {\bf \underline{if} } ( 
$A \setminus {\cal E}_{-,i} = \emptyset$
  ) {\bf  \underline{then} }\\
19. & $\qquad \quad$ {\bf \underline{return}} EXIT.\\
20. & $\qquad$ {\bf \underline{else} } \\
21. & $\qquad \quad$ Choose an edge $e$ from
$A \setminus {\cal E}_{-,i}$.\\
22. & $\qquad \quad$ $(e,\sigma,i)^{out} := (e,+,i)$.\\
23. & $\qquad \quad$ {\bf \underline{if} } ( $A \setminus {\cal E}_{-,i} = \{ e \}$  ) {\bf  \underline{then}}\\
24. & $\qquad \qquad$ {\bf \underline{return}} FIX.\\
25. & $\qquad \quad$ {\bf \underline{else}} \\
26. & $\qquad \qquad$  {\bf \underline{return}} BRANCH.\\
27. & $\qquad \quad$ {\bf \underline{end if}} \\
28. & $\qquad$ {\bf \underline{end if}} \\
29. & $\quad$ {\bf \underline{end if}} \\
30. & \\
31. & {\bf \underline{end for $i$} }\\
32. & \\
33. & {\bf \underline{return}} SUCCESS.\\
\end{tabular}
}}
\caption{Routine Packingclass\_test }
\label{pmctest}
\end{figure}     
Figure~\ref{pmctest} shows routine {\bf Packingclass\_test}.
As we have seen,
${\cal E}_{+}$ is a packing class, iff no excluded configuration
occurs in any coordinate direction. 
In this case, in each iteration of the $i$ loop, we keep
$A=\emptyset$, and the routine terminates in line 33.\ with
value $SUCCESS$.
Otherwise, $A$ contains a set of edges, out of which at least 
one has to be added to 
${\cal E}_{+,i}$ in order to remove the excluded configuration.
This edge must not be from ${\cal E}_{-,i}$.
Thus, the search on the subtree can be stopped for
$A \setminus {\cal E}_{-,i} = \emptyset$,
and
for $|A \setminus {\cal E}_{-,i}|=1$, the only edge
must be added to  ${\cal E}_{+,i}$.

Otherwise, an arbitrary edge from
$A \setminus {\cal E}_{-,i}$, together with a coordinate direction
$i$ is returned in the triple
$(e,\sigma,i)^{out}$ and used for branching.

In line 4.\ it is tested with the help of the decomposition
algorithm from \cite{GOL80} p.\ 129f.\ whether we have a comparability graph.
The runtime is
$O(\delta |E|)$, where $\delta$ is the maximal degree
of a vertex, and $E$ is the edge set of the examined graph.
It is simple to modify the algorithm, such that it returns
a $2$-chordless cycle in case of a negative result.

With the help of the algorithm from
\cite{GOL80} p.\ 133f., we can determine a maximal
weighted clique in a comparability graph in time that is linear
in the number of edges.
This algorithm is called in line 7., as graphs at this stage
have already passed the test for comparability graphs.

The search for a $C_{4}$ in line 10.\ can be realized by
two nested loops that enumerate possible pairs of opposite edges
in a potential $C_{4}$.

\subsection{Updating the Search Information}

\begin{figure} 
\fbox{\parbox{15.6cm}{
\begin{tabular}{ll}
{\bf Call: } &  {\bf Update\_$\,$searchinfo}($P$, ${(e,\sigma,i)}^{in}$, ${\cal E}$)\\
\\
{\bf Input: } & An OPP-$n$ instance $P:=(V,w)$, 
an augmentation 
$(e,\sigma,i)^{in}$, \\ 
& the search information ${\cal  E}$.\\
\\
{\bf Output: } &  
The updated search information ${\cal E}$, \\
&  and EXIT or OK. \\
\\
\end{tabular}

\begin{tabular}{rl}
1. & $(e,\sigma,i)$ := $(e,\sigma,i)^{in}$.\\
2. & \\
3. & {\bf \underline{if}} ( $(e,\sigma,i) = NULL$) {\bf \underline{then}}\\ 
4. & $\quad$ initialize ${\cal E}$ and mark this augmentation in $L$\\
5. & {\bf \underline{else}}\\ 
6. & $\quad$ {\bf \underline{if}} ( $\sigma = +$) {\bf \underline{then}}\\ 
7. & $\qquad$ ${\cal E}_{+,i} := {\cal E}_{+,i} \cup \{e\}$.\\
8. & $\qquad$ $L \quad := \{ (e,\sigma,i) \}$.\\
9. & $\quad$ {\bf \underline{else}}\\ 
10. & $\qquad$ {\bf \underline{for} } $f \in \clique{V}$ cannot be distinguished from $e$ {\bf \underline{do}}\\
11. & $\qquad$ $\quad$ ${\cal E}_{-,i} := {\cal E}_{-,i} \cup \{f\}$.\\
12. & $\qquad$ $\quad$ $L \quad := \{ (f,-,i) \}$.\\
13. & $\qquad$ {\bf \underline{end for $f$}}\\
14. & $\quad$ {\bf \underline{end if}}\\ 
15. & {\bf \underline{end if}}\\ 
16. & \\
17. & {\bf \underline{while}} ( $L \ne \emptyset$ ) {\bf \underline{do}}\\
18. & \\
19. & $\quad$ choose $(e,\sigma,i) \in L$.\\
20. & $\quad$ $L := L \setminus \{(e,\sigma,i)\}$.\\
21. & \\
22. & $\quad${\bf \underline{if}} ({\bf Check\_$\,$P3$\quad \!\!$ }($P$\,$(e,\sigma,i)$\,${\cal E}$\,$L$) $\ne$ OK ) 
{\bf \underline{then} \underline{return}} EXIT.\\
23. & $\quad${\bf \underline{if}} ({\bf Avoid\_C4$\qquad\!$ }($P$\,$(e,\sigma,i)$\,${\cal E}$\,$L$) $\ne$ OK ) 
{\bf \underline{then} \underline{return}} EXIT.\\
24. & $\quad${\bf \underline{if}} ({\bf Avoid\_cliques}($P$\,$(e,\sigma,i)$\,${\cal E}$\,$L$) $\ne$ OK ) {\bf \underline{then} \underline{return}} EXIT.\\
25. & \\
26. & {\bf \underline{end} \underline{while}}\\
27. & \\
28. & {\bf \underline{return}} OK.\\
\end{tabular}
}}
\caption{Routine Update\_$\,$searchinfo}
\label{fixierekanten}
\end{figure}     
Figure~\ref{fixierekanten} gives an overview over
Routine {\bf Update\_$\,$searchinfo}.
In the following, we will always refer to the current search node $N$
and denote the search information 
${\cal E}^{N}$ by ${\cal E}$.

The input triple $(e,\sigma,i)^{in}$ either describes
an augmentation of the search information
($e$ is fixed in ${\cal E}_{\sigma,i}$),
or it contains the value ``NULL'' on the root node.

On the root node, the search information is initialized as follows:
First the edges are fixed for which the vertices form
an infeasible stable set with two elements.
For $i \in \{1,\dots,d\}$, this means that all edges
$e=bc$ with $w_{i}(b)+w_{i}(c)>1$
are added to ${\cal E}_{+,i}$. 
Furthermore, we use Theorem~16 
from our paper \cite{pack2b}
(corresponding to Theorem~11 in \cite{pack2})
in order to fix cliques within the subgraphs
induced by the individual box types.

The augmentation 
$(e,\sigma,i)^{in}$ has either been fixed in the last branching step,
or it was returned by the routine
{\bf Packingclass\_test} together with the value ``FIX''.
In the latter case, 
$\sigma = +$ holds, so we know in case of $\sigma = -$
that the augmentation results from a branching step.
In Section~\ref{isosection} we concluded from
Lemma~\ref{vertauschlemma} that in this case,
all edges in ${\cal E}_{-,i}$ that are indistinguishable from $e$
can be fixed. This is done in lines 10.\ through 13.

In the main loop (lines 17. through 26.), for each augmentation of
${\cal E}$ it is checked whether it arises from a configuration
that allows it to fix further edges, or the search is stopped.
This recursive process is controlled by the list $L$.

The crucial work is done by the subroutines 
{\bf Check\_$\,$P3}, 
{\bf Avoid\_C4}, and
{\bf Avoid\_cliques}. 

\subsection*{Checking Condition $P3$}
In subroutine 
{\bf Check\_$\,$P3}, for an augmentation 
$(e,+,i)$ the set of ``free'' coordinate directions
$$                             
F:=\left\{ j \in \{1,\dots,d\} \, | \, e \notin {\cal E}_{+,j} \right\}
$$                             
is computed. If this set only has one element $k$, then for all
$E \in {\cal L}({\cal E})$ the condition $e \notin E_{k}$
must hold because of $\cap_{i=1}^d E_{i}=\emptyset$. 
In this case, $e$ can be fixed in ${\cal E}_{-,k}$,
and {\bf Check\_$\,$P3} terminates with the value ``OK''.
If there is no ``free'' coordinate direction left, then the search space
is empty, and the routine terminates with the value ``EXIT''.

\subsection*{Avoiding Induced $C_{4}$s}
\begin{figure} 
\fbox{\parbox{15.6cm}{
\begin{tabular}{ll}
{\bf Call: }& {\bf Avoid\_C4}($P$, $(e,\sigma,i)^{in}$, ${\cal E}$, $L$)\\
\\
{\bf Input: } & An OPP-$n$ instance $P:=(V,w)$, an augmentation
$(e,\sigma,i)^{in}$, \\
& the search information ${\cal E}$, the augmentation list $L$. \\
\\
{\bf Output: } &  
The updated search information ${\cal E}$, 
the updated \\
& augmentation list $L$, 
and the value EXIT or OK.\\
\\
\end{tabular}

\begin{tabular}{rl}
1. & $(e,\sigma,i)$ := $(e,\sigma,i)^{in}$.\\
2. & \\
3. & {\bf \underline{if}} ( $\sigma = +$ ) {\bf \underline{then}}\\
4. & \\
5. & $\quad$ {\bf \underline{for} } $f \notin {\cal E}_{-,i}$ completes a
$C_{4}$ in ${\cal E}_{+,i}$ \\ 
6. & $\qquad \quad$ 
that contains $e$ 
and has chords in ${\cal E}_{-,i}$. $\,\,$ {\bf \underline{do}}\\
7. & $\qquad$ {\bf \underline{if}} ( $f \in {\cal E}_{+,i}$ ) {\bf \underline{then} \underline{return}} EXIT.\\
8. & $\qquad$ ${\cal E}_{-,i} := {\cal E}_{-,i} \cup \{f\}$.\\
9. & $\qquad$ $L \quad := L \quad \cup \{(f,-,i)\}$.\\
10. & $\quad$ {\bf \underline{end for $f$} } \\
11. & \\
12. & $\quad$ {\bf \underline{for} } $f \notin {\cal E}_{+,i}$ is chord of a
$C_{4}$ in ${\cal E}_{+,i}$ \\ 
13. & $\qquad \quad$ 
that contains $e$ 
and has its other chord in ${\cal E}_{-,i}$. {\bf \underline{do}}\\
14. & $\qquad$ {\bf \underline{if}} ( $f \in {\cal E}_{-,i}$ ) {\bf \underline{then} \underline{return}} EXIT.\\
15. & $\qquad$ ${\cal E}_{+,i} := {\cal E}_{+,i} \cup \{f\}$.\\
16. & $\qquad$ $L \quad := L \quad \cup \{(f,+,i)\}$.\\
17. & $\quad$ {\bf \underline{end for $f$} } \\
18. & \\
19. & {\bf \underline{else}} ($\sigma = -$) \\
20. & \\
21. & $\quad$ {\bf \underline{for} } $f \notin {\cal E}_{-,i}$ completes a
$C_{4}$ in ${\cal E}_{+,i}$ that has $e$ as a chord\\ 
22. & $\qquad \quad$ 
and that has its other chord also in ${\cal E}_{-,i}$. $\,\,$ {\bf \underline{do}}\\
23. & $\qquad$ {\bf \underline{if}} ( $f \in {\cal E}_{+,i}$ ) {\bf \underline{then} \underline{return}} EXIT.\\
24. & $\qquad$ ${\cal E}_{-,i} := {\cal E}_{-,i} \cup \{f\}$.\\
25. & $\qquad$ $L \quad := L \quad \cup \{(f,-,i)\}$.\\
26. & $\quad$ {\bf \underline{end for $f$} } \\
27. & \\
28. & $\quad$ {\bf \underline{for} } $f \notin {\cal E}_{+,i}$ is chord of a
$C_{4}$ in ${\cal E}_{+,i}$ 
that has $e$ as its other chord. {\bf \underline{do}}\\
29. & $\qquad$ {\bf \underline{if}} ( $f \in {\cal E}_{-,i}$ ) {\bf \underline{then} \underline{return}} EXIT.\\
30. & $\qquad$ ${\cal E}_{+,i} := {\cal E}_{+,i} \cup \{f\}$.\\
31. & $\qquad$ $L \quad := L \quad \cup \{(f,+,i)\}$.\\
32. & $\quad$ {\bf \underline{end for $f$} } \\
33. & \\
34. & {\bf \underline{end if}}\\ 
35. & \\
36. & {\bf \underline{return}} OK.\\
\end{tabular}
}}
\caption{Routine Avoid\_C4}
\label{fixiereP1}
\end{figure}     

Routine {\bf Avoid\_C4} tries to detect edges
that can be used for completing an induced $C_{4}$
in $(V,{\cal E}_{+,i})$, with chords lying
in ${\cal E}_{-,i}$. 
Such an edge $f$ is then added to ${\cal E}_{+,i}$
or to ${\cal E}_{-,i}$, such that this excluded induced subgraph
is avoided.

Because this configuration must have been caused by the
augmentation $(e,\sigma,i)$ that was given to 
{\bf Avoid\_C4}, $e$ must either be an edge of the cycle,
or a chord. Because $f$ can occur as an edge of the cycle as well
as a chord, we have to check a total of four cases.
Figure~\ref{fixiereP1} shows the routine in detail.

\begin{figure} 
\fbox{\parbox{15.6cm}{
\begin{tabular}{ll}
{\bf Call: }& {\bf Avoid\_cliques}($P$, $(e,\sigma,i)^{in}$, ${\cal E}$, $L$)\\
\\
{\bf Input: } & An OPP-$n$ instance $(V,w)$, 
an augmentation 
$(e,\sigma,i)^{in}$, \\
& the search information ${\cal E}$, 
the augmentation list $L$. \\
\\
{\bf Output: } &  
The updated search information ${\cal E}$, 
the updated \\
& augmentation $L$, and EXIT or OK.\\
\\
\end{tabular}

\begin{tabular}{rl}
1. & $(bc,\sigma,i)$ := $(e,\sigma,i)^{in}$.\\
2. & \\
3. & {\bf \underline{if}} ( $\sigma = +$ ) 
{\bf \underline{then}}
{\bf \underline{return}} OK. \\
4. & \\
5. & compute $S'_{0}$ as described.\\
6. & {\bf \underline{if}} ( $w_{i}(S'_{0} \cup \{b,c\})>1$ ) 
{\bf \underline{then}}
{\bf \underline{return}} EXIT.  \\
7. & \\
8. & {\bf \underline{if}} ( $b$ and $c$ are indistinguishable ) {\bf \underline{then}}\\
9. & $\quad$ {\bf \underline{if} } $\exists b' \in V$ with $bb' \in \overline{{\cal E}_{+,i}} \cap \overline{{\cal E}_{-,i}}$ 
{\bf \underline{then} } \\
10. & $\qquad$ compute $B:=\{b,c\} \cup S' \cup X $ as described.\\
11. & $\qquad$ {\bf \underline{if}} ( $w_{i}(B) > 1$ ) {\bf \underline{then}}
${\cal E}_{+,i} := {\cal E}_{+,i} \cup \{bb'\}$, $L := L \cup \{(bb',+,i)\}$.\\
12. & $\quad$ {\bf \underline{end if} } \\
13. & {\bf \underline{end if} } \\
14. & \\
15. & {\bf \underline{for} } $b' \in V$ with $bb' \in \overline{{\cal E}_{+,i}} \cap \overline{{\cal E}_{-,i}}$ 
and $cb' \in {\cal E}_{-,i}$ 
{\bf \underline{do} } \\
16. & $\quad$ compute $B:=\{b,c\} \cup S' \cup X $ as described.\\
17. & $\quad$ {\bf \underline{if}} ( $w_{i}(B) > 1$ ) {\bf \underline{then}}
${\cal E}_{+,i} := {\cal E}_{+,i} \cup \{bb'\}$, $L := L \cup \{(bb',+,i)\}$.\\
18. & {\bf \underline{end for $b'$} } \\
19. & \\
20. & {\bf \underline{for} } $b' \in V$ with
$bb' \in {\cal E}_{-,i}$ 
and 
$cb' \in \overline{{\cal E}_{+,i}} \cap \overline{{\cal E}_{-,i}}$ 
{\bf \underline{do} } \\
21. & $\quad$ compute $B:=\{b,c\} \cup S' \cup X$ as described.\\
22. & $\quad$ {\bf \underline{if}} ( $w_{i}(B) > 1$ ) {\bf \underline{then}}
${\cal E}_{+,i} := {\cal E}_{+,i} \cup \{cb'\}$, $L := L \cup \{(cb',+,i)\}$.\\
23. & {\bf \underline{end for $b'$} } \\
24. & \\
25. & {\bf \underline{for} } $b' \in V$ with 
$bb' \in {\cal E}_{-,i}$ 
and 
$cb' \in {\cal E}_{-,i}$ 
{\bf \underline{do} } \\
26. & $\quad$ {\bf \underline{for} } $c' \in V$ with
$bc' \in {\cal E}_{-,i}$,    
$cc' \in {\cal E}_{-,i}$ and
$b'c' \in \overline{{\cal E}_{+,i}} \cap \overline{{\cal E}_{-,i}}$ 
{\bf \underline{do} } \\
27. & $\qquad$ compute $B:=\{b,c\} \cup S' \cup X $ as described.\\
28. & $\qquad$ {\bf \underline{if}} ( $w_{i}(B) > 1$ ) {\bf \underline{then}}
${\cal E}_{+,i} := {\cal E}_{+,i} \cup \{b'c'\}$, $L := L \cup \{(b'c',+,i)\}$.\\
29. & $\quad$ {\bf \underline{end for $c'$} } \\
30. & {\bf \underline{end for $b'$} } \\
31. & \\
32. & {\bf \underline{return}} OK.\\
\end{tabular}
}}
\caption{Routine Avoid\_cliques}
\label{fixiereP2}
\end{figure}     

\subsection*{Avoiding Infeasible Cliques}
The subroutine {\bf Avoid\_cliques} checks whether an
edge $e=bc$ that has been added to ${\cal E}_{-,i}$
completes one of the following configurations:
\begin{enumerate}
\item an $i$-infeasible clique
in $(V,{\cal E}_{-,i})$,
\item an $i$-infeasible clique
in $(V,\overline{{\cal E}_{+,i})}$, with edges not 
in 
${\cal E}_{-,i}$ being indistinguishable.
\end{enumerate}
As we have seen in
\ref{induzsubsection}, in the first case,
the search space is empty.
The routine terminates with value ``EXIT''.
In the second case, we can find a feasible augmentation by virtue
of Lemma~\ref{vertauschlemma}.

\bigskip
{\bf Computing $S_0'$:}

We search for a clique in
$(V,{\cal E}_{-,i})$ that contains $e=bc$
and has large weight.
Trivially, the box set of such a clique can only contain
$b$, $c$, and boxes from
$$
S_{0} := \{ z \in V \, | \, bz \in {\cal E}_{-,i} \, \wedge \, cz \in {\cal E}_{-,i} \}.
$$
Now our approach depends on whether
$(V,{\cal E}_{-,i})[S_{0}]$ is a comparability graph.
In the positive case, we can 
use the linear time algorithm from \cite{GOL80}
(just like for the test of packing classes)
to determine a set of boxes $S_{0}'$ that induces a maximal
weighted clique in 
$(V,{\cal E}_{-,i})[S_{0}]$.
Then $\{b,c\} \cup S_{0}'$ induces a clique in $(V,{\cal E}_{-,i})$
that has maximal weight among all cliques containing $e$.\\
If on the other hand, $(V,{\cal E}_{-,i})[S_{0}]$ is not a comparability
graph, then we skip the computation of a maximal weighted clique.
(As a generalization of the CLIQUE problem (problem [GT19] in \cite{GAJO79}),
this problem is NP-hard.) Instead, we compute
$S'_{0}$ by using a greedy strategy.
Starting with $S'_{0}=\emptyset$, we keep augmenting
$S'_{0}$ by the box with the largest
weight $w_{i}$, as long as the property $\clique{S_0'} \subseteq {\cal E}_{-,i}$
remains valid.
The clique induced by 
$\{b,c\} \cup S'_{0}$ in $(V,{\cal E}_{-,i})$ may be suboptimal.\\
In both cases, the routine terminates
in case of an $i$-infeasible
$S'_{0} \cup \{b,c\}$ with the value ``EXIT''.

\bigskip
In order to determine whether 
$(V,{\cal E}_{-,i})[S_{0}]$ is a comparability graph,
we use the decomposition algorithm
from \cite{GOL80} in {\bf Packingclass\_test}.
In the implementation,
it is worthwhile taking into account that in the case
$|S_{0}| \le 4$, the testing for a comparability graph can be omitted.
The corresponding induced subgraphs must be comparability graphs,
because a $2$-chordless cycle must contain at least five different vertices.
In our numerical experiments, this turned out to be a common situation.

\bigskip
{\bf Finding an augmenting edge by computing $B$:}

We test whether an edge 
$e' \in \overline{{\cal E}_{+,i}} \cap \overline{{\cal E}_{-,i}}$ can be fixed.
A sufficient condition is the existence of a set 
$B \subseteq V$ that satisfies the following conditions:
\begin{enumerate}
\item $B$ contains all vertices of $e$ and $e'$.
\item All edges in $\clique{B} \setminus {\cal E}_{-,i}$ 
are indistinguishable.
\item $B$ is $i$-infeasible.
\end{enumerate}
Because of $P2$, an edge in $\clique{B}$ must be in the $i$th component graph
of the desired packing class. Because this edge must not be in
${\cal E}_{-,i}$, it must be indistinguishable from
$e' \in \clique{B} \setminus {\cal E}_{-,i}$ by virtue of 2.
In other words, Lemma~\ref{vertauschlemma} 
means that augmentation with indistinguishable edges leads to
isomorphic packing classes. This implies the feasibility
of augmentation $(e',+,i)$.

The requirement that the vertices of $e$ lie in $B$
results from the fact that we only search for incomplete excluded 
configurations that arise from adding $e$ to ${\cal E}_{-,i}$.

\begin{figure}
\begin{center}
\leavevmode
\epsfxsize=10cm
\epsffile{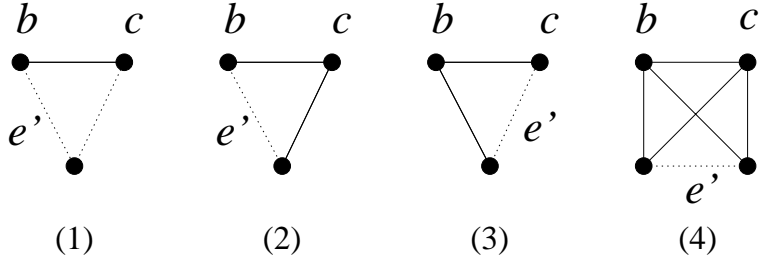}
\caption{Relative position of $e'$ and $e=bc$.}
\label{vierP2pic}
\end{center}
\end{figure}

When identifying edges that are candidates for $e'$, we get four cases
for the position of $e'$ relative to $e=bc$
in $\clique{B}$, as shown in Figure~\ref{vierP2pic}.
Dashed lines represent the (unfixed) edges 
in $\overline{{\cal E}_{+,i}}\cap\overline{{\cal E}_{-,i}}$,
while solid lines represent edges in ${\cal E}_{-,i}$.
The second requirement for $B$ implies that $b$ and $c$ are indistinguishable.
Note that after the first resulting augmentation, $b$ and $c$
are indistinguishable with respect to the current search information.
Thus, the cases (1), (2), (3), and (4) in the figure correspond
to lines (8.--13.), (15.--18.), (20.--23.), and (25.--30.): 

Therefore, constructing the set $B$ for a given edge $e'$ is done as follows.
Let $S$ be the set of boxes that are adjacent in $(V,{\cal E}_{-,i})$
to all vertices of $e$ and $e'$.
Similar to the above construction of $S_{0}'$ from $S_{0}$,
we construct a set of boxes $S'$ from $S$ that induces a clique
in $(V,{\cal E}_{-,i})$.
The comparability graph test is skipped, if 
$(V,{\cal E}_{-,i})[S_{0}]$ has been recognized as a comparability graph:
If $S \subseteq S_{0}$, then
$(V,{\cal E}_{-,i})[S]$ is an induced subgraph
and inherits its property of being a comparability graph.

By adding the vertices of $e$ and $e'$ to $S'$,
we get a set that satisfies the first two conditions that $B$ needs to satisfy.
$e'$ is the only set in the complete graph on this set
that does not belong to ${\cal E}_{-,i}$.
Now we add boxes that provide edges indistinguishable from $e'$.

Let the set $X$ contain the vertices of $e'$.
The indistinguishable boxes for each vertex form a stable set
in $(V,{\cal E}_{-,i})$, or they induce a clique in this graph.
Only in the latter case do we add these boxes to $X$.
With the help of this construction, any edge in the complete
graph on
$B:=\{b,c\} \cup S' \cup X$ is either in the set
${\cal E}_{-,i}$, or it is indistinguishable from $e'$.
If this set is $i$-infeasible, then we fix
$e'$ in ${\cal E}_{+,i}$ by virtue of Lemma~\ref{vertauschlemma}.
 
\section{A Tree Search Algorithm for Orthogonal Knapsack Problems} 
\label{stufe1sec}

In this section, we elaborate how the data structure introduced 
in the paper~\cite{pack1b,pack1}, the lower bounds described in~\cite{pack2b,pack2},
and the exact algorithm for the OPP from Section~\ref{oppsec}
can be used as building blocks for new exact methods for
orthogonal packing problems.

We concentrate on the most difficult problem, the OKP.
After a detailed description of the new branch-and-bound approach,
the following Section~\ref{comput}
gives evidence that our algorithm
allows it to solve considerably larger
instances than previous methods. In particular, we present
the first results for 3-dimensional instances.

Similar exact algorithms for the SPP and the OBPP are sketched 
in Section~\ref{other}.

\subsection{The Framework} \label{bbsec}
For solving the OKP, 
we have to determine a subset
$S \subseteq V$ of boxes that has maximum value among all subsets
of boxes fitting into the container.
Like Beasley~\cite{BEA85}, and Christofides/Hadjiconstantinou~\cite{HACHR95},
we will prove feasibility of a particular set $S$ 
by displaying a feasible packing for $(V,w,W)$.
For most practical applications, this is of key importance.

In the branch-and-bound algorithms \cite{BEA85} and \cite{HACHR95},
the iterative choice of a subset and the corresponding
packing are treated simultaneously:
With each branching step, it is decided whether a particular
position in the container is occupied by a particular box type.

In contrast to this, our approach works in two levels.
Only after the first level has determined the subset
$S \subseteq V$ will the OPP algorithm from Section~\ref{oppsec}
try to find a feasible packing.
This allows us to use the lower bounds described 
in our paper~\cite{pack2b,pack2}
for excluding most of the first level search tree without having to
consider the particular structure of a packing.
Our numerical results show that only in a small fraction of search nodes,
the second level search has to be used. Note that the main innovation
of our approach lies in this second level; it is to be expected that tuning
the outer level (as was done by Caprara and Monaci~\cite{CM04})
yields even better results.

\subsection{Branch-and-Bound Methods}
We assume that the reader is familiar with the general structure
of a branch-and-bound algorithm. (A good description can be 
found in~\cite{NEWO88}.) We start by introducing some notation.

We remind the reader of the partitioning of the set $V$ of boxes
into classes of boxes with identical size and value, called {\em box types}:
\[
V = \parti_{t=1}^{m} T_{t}.
\]
For box type $t$, we set 
$n_{t} := |T_{t}|$ and denote the elements by
\[
T_{t} =: \{ b_{t,1}, \dots, b_{t,n_{t}} \}.
\] 
For ease of notation, we write
$w^{(t)}$ instead of $w(b_{t,1})$, and
$v^{(t)}$ instead of $v(b_{t,1})$.

In the test instances that we will be dealing with,
all sizes of boxes and containers are integers.
We denote measures of the container by
$W \in \nat^{d}$;  when discussing mathematical arguments,
we will assume without loss of generality that the container is a unit
cube. 

\subsection{Search Nodes at Level One}

The first-level search tree enumerates the subsets
$S \subseteq V$ that are candidates for a solution subset for the OKP.
Each node $N$ of the search tree corresponds to
an OKP instances with the additional constraint
that each box type
$T_{t}$ has upper and lower bounds for the number of boxes that are used.
These bounds are denoted by 
$\overline{n}_{t}^{N}$ and $\underline{n}_{t}^{N}$.

For a search node $N$, we set  
$$
\underline{S}^{N} := \parti_{t=1}^{m} \{ b_{t,1}, \dots, b_{t,\underline{n}^{N}_{t}} \} 
$$
and similarly 
$$
\overline{S}^{N} := \parti_{t=1}^{m} \{ b_{t,1}, \dots, b_{t,\overline{n}^{N}_{t}} \}.
$$
For a partial search tree with root $N$, only subsets $S$ will be considered
that satisfy
$$
{\cal S}(N) := \{ S \subseteq V \, | \, \underline{S}^{N} \subseteq S \subseteq \overline{S}^{N}. \}
$$
Thus, for a search node $N$, the corresponding restricted OKP
is given by
\begin{equation} \label{restokp}     
\begin{tabular}{rl}
{\bf Maximize }& $v(S)$, \\ [0.4ex]
{\bf such that}& there is a feasible packing for $(S,w)$, \\
    & $\underline{S}^{N} \subseteq S \subseteq \overline{S}^{N}$. 
\end{tabular}
\end{equation}                   

On the root node $N_{0}$, we start with the original problem, i.\ e.,
$\underline{n}_{t}^{N_{0}}=0$ and
$\overline{n}_{t}^{N_{0}}=n_{t}$ for $t \in \{1,\dots,m\}$.
Then $\underline{S}^{N_{0}} = \emptyset$ and
$\overline{S}^{N_{0}} = V$.

Enumerating the first level search tree is done by
{\em best first search}:
Each node $N$ is assigned a preliminary local upper bound, given by
the minimum of $v(\overline{S}^{N})$ and the local upper bound
of its parent node. (A better upper bound will be determined
while evaluating the partial tree at $N$.)
At each stage, we choose a new node where this local upper bound is
maximal.

\subsection{Branching}
When a subset $S$ has been uniquely determined by 
the condition $S \in {\cal S}(N)$ ,
we have reached a leaf of the first level search tree.
In this case, we have
$$
\underline{S}^{N} = S = \overline{S}^{N}
$$
and 
$$
\forall t  \in \{1,\dots,m\}: \quad
\underline{n}_{t}^{N} = \overline{n}_{t}^{N}.
$$
Then the problem
(\ref{restokp}) is an OPP that is solved by the second level tree search.

Otherwise, we have box types $T_{t}$, with
$\underline{n}_{t}^{N} < \overline{n}_{t}^{N}$.
We choose the one with largest size 
$\max_{1 \le i \le d} w_{i}^{(t)}$
for an arbitrary coordinate direction.
By our experience, boxes that are ``bulky'' in this sense
have the biggest influence on the overall solution of the problem.

Now let $T_{t^{\ast}}$ be the box type chosen in this way.
We branch by splitting ${\cal S}(N)$ 
into subspaces, where the number of boxes in
$T_{t^{\ast}}$ is constant.
For each
$\nu \in \{\underline{n}_{t^{\ast}}^{N}, \dots, \overline{n}_{t^{\ast}}^{N} \},$
we determine a child node 
$N_{\nu}$.
For this node, we set
$$
\underline{n}_{t}^{N_{\nu}} := \left\{ \begin{array}{ll} \nu, & t=t^{\ast}, \\ 
\underline{n}_{t}^{N} & t \ne t^{\ast}, \end{array} \right. 
$$
and
$$
\overline{n}_{t}^{N_{\nu}} := \left\{ \begin{array}{ll} \nu, & t=t^{\ast}, \\ 
\overline{n}_{t}^{N} & t \ne t^{\ast}. \end{array} \right.
$$
A different branching strategy builds a binary search tree,
where the two children of $N$ each get one half of
$\{\underline{n}_{t^{\ast}}^{N}, \dots, \overline{n}_{t^{\ast}}^{N} \}$ 
as a range for the number of boxes in 
$T_{t^{\ast}}$.
For technical reasons, we have used the first variant.

\subsection{Lower Bounds}

On each node $N$, the container is filled with boxes from $\overline{S}^{N}$
by using the following greedy heuristic.
The best objective value of the OKP for any of these feasible solutions
is stored in $v_{lb}$.
 Trivially, $v_{lb}$ is a lower bound for the optimal value of the OKP.

In our heuristic, we build a sequence of packings, where each
lower coordinate of a box equals $0$ (i.\ e., the boundary of the container),
or the upper coordinate of a preceding box.
These positions are called 
{\em placement points}. Placement points are maintained in a list 
that is initialized by the container origin. 
At each step, a placement point is removed from the list,
as we try to use it for placing another box. Following a given ordering,
we use the first box type that fits at the chosen placement point without
overlapping any of the boxes that are already packed into the container.
In case of success, we compute the new placement points and add them to the
list. This step is repeated until the list is empty, or all boxes
have been packed.

This construction of a packing is repeated for several orderings
of box types. In the first round, we use the order of decreasing value.
Following rounds use a random weighting of values before sorting;
weights are chosen from a uniform distribution on $[0,1]$.

In our implementation, 50 iterations of this heuristic are performed at the
root, and 10 iterations at all other nodes.

\subsection{Upper Bounds}

The upper bound
$v_{ub}^{N}$ refers to the set of boxes from
${\cal S}(N)$. 
As we showed in the paper~\cite{pack2b,pack2}, for any conservative
scale $w'$ for $(\overline{S}^{N},w)$, a relaxation of
(\ref{restokp}) is given by
\begin{equation} \label{restrel}   
\begin{tabular}{rl}
{\bf Maximize }& $v(S)$,\\ [0.4ex]
{\bf such that }& 
$\sum_{b\in S}\otimes w'(b) \le 1$, \\
& $\underline{S}^{N} \subseteq S \subseteq \overline{S}^{N}$,     
\end{tabular}
\end{equation}                   
where $\otimes w'(b):= \prod_{i=1}^d w'_i(b)$ denotes
the volume of the modified box $w'(b)$.
In order to avoid technical difficulties, we only consider
conservative scales that are constant for each box type.
For the benefit of later generalizations, we formulate the problem
(\ref{restrel}) explicitly as a restricted one-dimensional knapsack problem:
\begin{eqnarray} \label{okprestrelax}
\mbox{\bf Maximize }&& \sum_{t=1}^{m} v(b_{t,1}) \xi_{t},\\
\mbox{\bf such that } && \sum_{t=1}^{m} \otimes w'(b_{t,1}) \xi_{t} \le 1, \nonumber \\
 && \underline{n} \le \xi \le \overline{n}, \nonumber \\
 && \xi \mbox{ integer} \nonumber.
\end{eqnarray}
A problem of this type can be solved by the routine
Routine MTB2 from~\cite{MATO90}, Appendix A.3.1.
This transforms the restricted knapsack problem into a
0-1 knapsack problem to which the algorithm 
of Martello and Toth (\cite{MATO90}, pp. 61ff.) is applied.

In our implementation, we use as an upper bound the minimum of the optimal
values of the relaxations (\ref{restrel}) 
for the conservative scales 
$$
w' \in \{(w_{1}, \dots, u^{(k)} \circ w_{i}, \dots, w_{d}) \, | \, i = 1,\dots,d, \quad
k = 1,2,3,4 \}
$$
from our paper~\cite{pack2b,pack2}.

\subsection{Removal of Partial Search Trees}
We can stop the search on the current search tree, if one of the following
conditions is satisfied:
\begin{enumerate}
\item $v_{ub}^{N} \le v_{lb}$.
\item $\overline{S}^{N}$ fits into the container.      
\item $\underline{S}^{N}$ does not fit into the container. 
\end{enumerate}
The first stop criterion is used in any branch-and-bound procedure.
In this case, the currently best solution cannot be improved on the
current search tree.

In the second case,
$\overline{S}^{N} \in {\cal S}(N)$ is a best feasible solution in
${\cal S}(N)$. Because we are always trying to pack all of $\overline{S}^{N}$
when updating the lower bound $v_{lb}$, this condition is checked 
when performing the update.

In the third case, 
$\underline{S}^{N} \subseteq S$ implies that no set $S \in {\cal S}(N)$
can be packed into the container, so
${\cal S}(N)$ cannot contain a feasible solution.
This means we have to solve another OPP.

\subsection{Solving Orthogonal Packing Problems}
In order to solve the OPPs that occur on the leaves of
the search tree and when checking the stop criterion
``$\underline{S}^{N}$ does not fit into the container'',
we use the following strategy: 

First we try to use the volume criterion for a selection of
conservative scales. Other than the original weight function $w$,
we use the conservative scales
$$
w' \in \{(w_{1}, \dots, u^{(k)} \circ w_{i}, \dots, w_{d}) \, | \, i = 1,\dots,d, \quad
k = 1,\dots,W_{i}/2 \}.
$$
If this does not produce a (negative) result, we try to find a packing pattern
by 10 iterations of our search heuristic.
If this fails as well, we use our algorithm from Section~\ref{oppsec}
to decide the OPP.

\subsection{Problem Reduction}
There are several ways to decrease the gap between the bounds 
$\underline{n}_{t}^{N}$ and $\overline{n}_{t}^{N}$ on a search node.
These rules are based upon corresponding reduction tests
of Beasley~\cite{BEA85}. If the areas of boxes and container are used,
we generalize the tests from two to $d$ dimensions.
By using conservative scales, we get a generalization of these
tests, with markedly increased efficiency.

We start with the rule
{\em Free Value} that remains unchanged.
An optimal solution $S$ can have at most value 
$v_{ub}^{N}$. Because $\underline{S} \subseteq S$,  
further boxes from $T_{t}$ in $S$ can contribute at most a value of
$v_{ub}^{N} - v(\underline{S})$. 
Because each of these boxes has value $v^{(t)}$, we set
\begin{equation} \label{freevalue}
\overline{n}_{t} := \min \left\{ \overline{n}_{t}, \underline{n}_{t} + \left\lfloor \frac{ v_{ub}^{N} - 
 v(\underline{S}) }{ v^{(t)} } \right\rfloor \right\}.
\end{equation}

A similar argument, used on volumes, is the basis for Beasley's reduction test
{\em Free Area}.
The volume used by $\underline{S}$ is at least as big as the sum
of the volumes of the individual boxes in 
$\underline{S}$. Further boxes from $T_{t}$ can use
at most the volume of the container, reduced by this amount.
Because each of these boxes uses a volume
of $\otimes w{(t)}$, we can use the following update for 
$t \in \{1,\dots,m\}$:
\begin{equation} \label{freearea}
\overline{n}_{t} := \min \left\{ \overline{n}_{t}, \underline{n}_{t} + \left\lfloor \frac{ 1 - 
 \otimes w(\underline{S}) }{ \otimes w({t}) } \right\rfloor. \right\}
\end{equation}
With the help of Corollary~8 
from our paper \cite{pack2b}, 
we can generalize
{\em Free Area} by replacing $w$ in
(\ref{freearea}) by an arbitrary conservative scale
$w'$ for $(V,w)$ that is constant on $T_{t}$.
In our implementation, we use $w$, and the conservative scales
$$
w' \in \{(w_{1}, \dots, u^{(k)} \circ w_{i}, \dots, w_{d}) \, | \, i = 1,\dots,d, \quad
k = 1,\dots,W_{i}/2 \}.$$

To allow for further possible improvement of a bound
$\underline{n}_{t^{\ast}}^{N}$, $t^{\ast} \in \{1,\dots,m\}$, 
we expand the relaxation (\ref{okprestrelax}) by the additional
constraint $\xi_{t^{\ast}} = \underline{n}_{t^{\ast}}$. The optimal values
of the resulting knapsack problems are upper bounds for the value
of those solutions $S \in {\cal S}(N)$ that contain
precisely 
$\underline{n}_{t^{\ast}}$ boxes from $T_{t^{\ast}}$.
If the minimum of these bounds does not exceed
$v_{lb}$, a solution $S \in {\cal S}(N)$ with a better value
than the current best must contain more than
$\underline{n}_{t^{\ast}}^{N}$                                
boxes from $T_{t^{\ast}}$.
In this case, we can increment $\underline{n}_{t^{\ast}}^{N}$.                                
This test is repeated for each box type $t$, until no bound can be improved.
If in this process, we get $\underline{n}_{t}^{N} > \overline{n}_{t}^{N}$,
then the search on the partial search tree with root $N$ can be
stopped. Thus, we have derived a generalization of 
the reduction test {\em Area Program} with the help of conservative scales.
 
\section{Computational Results}
\label{comput}
The above OKP procedure has been implemented in C++
and was tested on a Sun workstation with Ultra SPARC processors (175 MHz),
using the compiler gcc. To allow for a wider range of comparisons with other
two-dimensional efforts, we also tested an implementation 
on a PC with Pentium IV processor (2,8 GHz) with 1 GB memory using g++3.2.

\subsection{Results for Benchmark Instances from the Literature}
\begin{table}[hp]
{\scriptsize
\begin{center}
\leavevmode

\begin{tabular}{|l|r|r|r|r|r|r|r|r|}
\hline
problem   & container    &  box  & \#    & OKP   & OPP   &  OPP  & opt.  & opt. \\
          & size         & types & boxes & nodes & calls &  nodes & boxes & sol. \\
\hline
beasley1  & (  10,  10) &     5 &    10 &    19 &     1 &      1 &      5 &    164 \\ 
beasley2  & (  10,  10) &     7 &    17 &     5 &     0 &      0 &      5 &    230 \\ 
beasley3  & (  10,  10) &    10 &    21 &    25 &     6 &     36 &      7 &    247 \\ 
beasley4  & (  15,  10) &     5 &     7 &     1 &     0 &      0 &      6 &    268 \\ 
beasley5  & (  15,  10) &     7 &    14 &     1 &     0 &      0 &      6 &    358 \\ 
beasley6  & (  15,  10) &    10 &    15 &    15 &     5 &      5 &      7 &    289 \\ 
beasley7  & (  20,  20) &     5 &     8 &     0 &     0 &      0 &      8 &    430 \\ 
beasley8  & (  20,  20) &     7 &    13 &    53 &    23 &    301 &      8 &    834 \\ 
beasley9  & (  20,  20) &    10 &    18 &     3 &     0 &      0 &     11 &    924 \\ 
beasley10 & (  30,  30) &     5 &    13 &     1 &     0 &      0 &      6 &   1452 \\ 
beasley11 & (  30,  30) &     7 &    15 &    36 &    10 &     16 &      9 &   1688 \\ 
beasley12 & (  30,  30) &    10 &    22 &    48 &    14 &    105 &      9 &   1865 \\ 
\hline                                                          
hadchr3   & (  30,  30) &     7 &     7 &     1 &     0 &      0 &      5 &   1178 \\ 
hadchr7   & (  30,  30) &    10 &    22 &    48 &    14 &    105 &      9 &   1865 \\ 
hadchr8   & (  40,  40) &    10 &    10 &     7 &     0 &      0 &      6 &   2517 \\ 
hadchr11  & (  30,  30) &    15 &    15 &    30 &     1 &      1 &      5 &   1270 \\ 
hadchr12  & (  40,  40) &    15 &    15 &     5 &     0 &      0 &      7 &   2949 \\ 
\hline                                                          
wang20    & (  70,  40) &    20 &    42 &   794 &   176 &   1003 &      8 &   2726 \\ 
chrwhi62  & (  40,  70) &    20 &    62 &   356 &   102 &   7991 &     10 &   1860 \\ 
\hline                                                          
3         & (  40,  70) &    20 &    62 &   356 &   102 &   7991 &     10 &   1860 \\ 
3s        & (  40,  70) &    20 &    62 &   757 &   166 &   3050 &      8 &   2726 \\ 
A1        & (  50,  60) &    20 &    62 &   935 &   254 &  19283 &     11 &   2020 \\ 
A1s       & (  50,  60) &    20 &    62 &  4291 &   504 &   8156 &      7 &   2956 \\ 
A2        & (  60,  60) &    20 &    53 &   267 &    70 &  35747 &     11 &   2615 \\ 
A2s       & (  60,  60) &    20 &    53 &  8598 &  2365 & 143002 &      8 &   3535 \\ 
CHL2      & (  62,  55) &    10 &    19 &   688 &   317 & 225011 &      9 &   2326 \\ 
CHL2s     & (  62,  55) &    10 &    19 &  1419 &   557 & 158450 &     10 &   3336 \\ 
CHL3      & ( 157, 121) &    15 &    35 &     0 &     0 &      0 &     35 &   5283 \\ 
CHL3s     & ( 157, 121) &    15 &    35 &     0 &     0 &      0 &     35 &   7402 \\ 
CHL4      & ( 207, 231) &    15 &    27 &     0 &     0 &      0 &     27 &   8998 \\ 
CHL4s     & ( 207, 231) &    15 &    27 &     0 &     0 &      0 &     27 &  13932 \\ 
CHL5      & (  30,  20) &    10 &    18 &   363 &   194 &  57115 &     11 &    589 \\ 
\hline                                                          
cgcut1    & (  15,  10) &     7 &    16 &    14 &     1 &      1 &      8 &    244 \\ 
cgcut2    & (  40,  70) &    10 &    23 &       &       &        &     12 &   2892 \\
cgcut3    & (  40,  70) &    20 &    62 &   356 &   102 &   7991 &     10 &   1860 \\ 
\hline                                                          
gcut01    & ( 250, 250) &    10 &    10 &    33 &     0 &      0 &      3 &  48368 \\ 
gcut02    & ( 250, 250) &    20 &    20 &   519 &    51 &     78 &      6 &  59798 \\ 
gcut03    & ( 250, 250) &    30 &    30 &  2234 &   235 &    742 &      6 &  61275 \\ 
gcut04    & ( 250, 250) &    50 &    50 & 72159 & 18316 & 145057 &      4 &  61380 \\ 
gcut05    & ( 500, 500) &    10 &    10 &    52 &    13 &     13 &      5 & 195582 \\ 
gcut06    & ( 500, 500) &    20 &    20 &   278 &    22 &     22 &      4 & 236305 \\ 
gcut07    & ( 500, 500) &    30 &    30 &   852 &   124 &    152 &      4 & 240143 \\ 
gcut08    & ( 500, 500) &    50 &    50 & 55485 &  9037 &  15970 &      4 & 245758 \\ 
gcut09    & (1000,1000) &    10 &    10 &    12 &     2 &      8 &      5 & 939600 \\ 
gcut10    & (1000,1000) &    20 &    20 &   335 &    31 &     40 &      5 & 937349 \\ 
gcut11    & (1000,1000) &    30 &    30 &  1616 &   212 &    463 &      6 & 969709 \\ 
gcut12    & (1000,1000) &    50 &    50 &  8178 &   593 &   1236 &      5 & 979521 \\ 
gcut13    & (3000,3000) &    32 &    32 &       &       &        &        &$\ge$8622498\\
          &             &       &       &       &       &        &        &$\le$9000000\\
\hline                                                          
okp1      & ( 100, 100) &    15 &    50 &  3244 &   661 &  35523 &     11 &  27718 \\ 
okp2      & ( 100, 100) &    30 &    30 & 23626 &  7310 &   8721 &     11 &  22502 \\ 
okp3      & ( 100, 100) &    30 &    30 &  8233 &   816 &    921 &     11 &  24019 \\ 
okp4      & ( 100, 100) &    33 &    61 &  1458 &    15 &     50 &     10 &  32893 \\ 
okp5      & ( 100, 100) &    29 &    97 &  5733 &   643 &  13600 &      8 &  27923 \\ 
\hline
\end{tabular}
\caption{Two-dimensional benchmark instances from previous literature.}
\label{problems}
\end{center}
}

\end{table}
\begin{table}[hp]
{\scriptsize
\begin{center}
\leavevmode

\begin{tabular}{|l|r|r|r|r|r|r|r|r|r|r|r|}
\hline
problem    & time/s &  B85  & HC95   & \multicolumn{4}{|c|}{CM04} \\
           &        &       &        & $A_0$    & $A_1$    & $A_2$    & $A_3$          \\
\hline						                                 
beasley1   &$<$0.01 &   0.9 &        &          &          &          &              \\
beasley2   &$<$0.01 &   4.0 &        &          &          &          &              \\
beasley3   &$<$0.01 &  10.5 &        &          &          &          &              \\
beasley4   &$<$0.01 &   0.1 &   0.04 &          &          &          &              \\
beasley5   &$<$0.01 &   0.4 &        &          &          &          &              \\
beasley6   &$<$0.01 &  55.2 &  45.20 &          &          &          &              \\
beasley7   &$<$0.01 &   0.5 &   0.04 &          &          &          &              \\
beasley8   &   0.02 & 218.6 &        &          &          &          &              \\
beasley9   &$<$0.01 &  18.3 &   5.20 &          &          &          &              \\
beasley10  &$<$0.01 &   0.9 &        &          &          &          &              \\
beasley11  &$<$0.01 &  79.1 &        &          &          &          &              \\
beasley12  &   0.02 & 229.0 & $>$800 &          &          &          &              \\
\hline               				                                 
hadchr3    &$<$0.01 &       &    532 &          &          &          &              \\
hadchr7    &   0.01 &       & $>$800 &          &          &          &              \\
hadchr8    &$<$0.01 &       & $>$800 &          &          &          &              \\
hadchr11   &$<$0.01 &       & $>$800 &          &          &          &              \\
hadchr12   &$<$0.01 &       &   65.2 &          &          &          &              \\
\hline               				                                 
wang20     &   0.67 &       &        &     6.75 &     6.31 &    17.84 &     2.72 \\
chrwhi62   &   0.54 &       &        &          &          &          &          \\
\hline               				                                 
3          &   0.54 &       &        &          &          &          &          \\
3s         &   0.46 &       &        &          &          &          &          \\
A1         &   1.12 &       &        &          &          &          &          \\
A1s        &   1.51 &       &        &          &          &          &          \\
A2         &   1.62 &       &        &          &          &          &          \\
A2s        &   8.35 &       &        &          &          &          &          \\
CHL2       &  10.36 &       &        &          &          &          &          \\
CHL2s      &   6.84 &       &        &          &          &          &          \\
CHL3       &$<$0.01 &       &        &          &          &          &          \\
CHL3s      &$<$0.01 &       &        &          &          &          &          \\
CHL4       &$<$0.01 &       &        &          &          &          &          \\
CHL4s      &$<$0.01 &       &        &          &          &          &          \\
CHL5       &   4.66 &       &        &          &          &          &          \\
\hline               				                                 
cgcut1     &$<$0.01 &       &        &     0.30 &     1.47 &     1.46 &     1.46 \\
cgcut2     &$>$1800 &       &        &  $>$1800 &  $>$1800 &   533.45 &   531.93 \\
cgcut3     &   0.54 &       &        &    23.76 &    23.68 &     4.59 &     4.58 \\
\hline               				                                 
gcut1      &   0.01 &       &        &     0.00 &     0.00 &     0.01 &     0.01 \\
gcut2      &   0.47 &       &        &     0.52 &     0.19 &    25.75 &     0.22 \\
gcut3      &   4.34 &       &        &  $>$1800 &     2.16 &   276.37 &     3.24 \\
gcut4      & 195.62 &       &        &  $>$1800 &   346.99 &  $>$1800 &   376.52 \\
gcut5      &   0.02 &       &        &     0.00 &     0.50 &     0.03 &     0.50 \\
gcut6      &   0.38 &       &        &     0.06 &     0.09 &     9.71 &     0.12 \\
gcut7      &   2.24 &       &        &     1.31 &     0.63 &   354.50 &     1.07 \\
gcut8      & 253.54 &       &        &  1202.09 &   136.71 &  $>$1800 &   168.50 \\
gcut9      &   0.01 &       &        &     0.01 &     0.09 &     0.05 &     0.08 \\
gcut10     &   0.67 &       &        &     0.01 &     0.13 &     6.49 &     0.14 \\
gcut11     &   8.82 &       &        &    16.72 &    14.76 &  $>$1800 &    16.30 \\
gcut12     & 109.81 &       &        &    63.45 &    16.85 &  $>$1800 &    25.39 \\
gcut13     &$>$1800 &       &        &  $>$1800 &  $>$1800 &  $>$1800 &  $>$1800 \\
\hline               				                                 
okp1       &  10.82 &       &        &    24.06 &    25.46 &    72.20 &    35.84 \\
okp2       &  20.25 &       &        &  $>$1800 &  $>$1800 &  1535.95 &  1559.00 \\
okp3       &   5.98 &       &        &    21.36 &     1.91 &   465.57 &    10.63 \\
okp4       &   2.87 &       &        &    40.40 &     2.13 &     0.85 &     4.05 \\
okp5       &  11.78 &       &        &    40.14 &  $>$1800 &   513.06 &   488.27 \\
\hline
\end{tabular}
\caption{Runtimes of our implementation, compared to other methods.
The columns ``B85'', ``HC95'', ``CM04''
show the runtimes as reported in \cite{BEA85,HACHR95,CM04}.}
\label{results}
\end{center}
}
\end{table}

The only
benchmark instances for the OKP that have been documented in the literature
can be found in the articles by 
Beasley~\cite{BEA85}, and Hadjiconstantinou and Christofides~\cite{HACHR95}.
These are restricted to the two-dimensional case. We ran our algorithm
on all of these instances that were available.
Like Caprara and Monaci~\cite{CM04}, we also use a number of other
instances that were originally designed for guillotine-cut instances.

The twelve instances {\em beasley1} through {\em beasley12} are taken 
from Beasley's OR library (see~\cite{ORLIB90}). 
They can be found on the internet at 
$$
\mbox{\tt http://mscmga.ms.ic.ac.uk/jeb/orlib/ngcutinfo.html} 
$$
The data for 
{\em hadchr3} and {\em hadchr11} is given in~\cite{HACHR95}.

Tables~\ref{problems} and \ref{results} show our results for these
OKP-$2$ instances. For all instances, we found an optimal solution
in at most 0.02 seconds. The small number of OKP search nodes
(at most 65) as well as OPP search nodes (at most 294) shows the
high efficiency of the rules for reducing the search tree.
It is also remarkable that on less than a quarter of the search nodes,
the enumeration procedure for the OPP had to be used.
The majority of reductions resulted from transformed volumes.

Instances {\em wang20} and {\em chrwhi62} are considerably larger.
They were taken from
Wang~\cite{WANG83}, and Christofides and Whitlock~\cite{CHWH77} 
and originally designed for testing the efficiency of
guillotine cut algorithms, as were the next three sets:
Instances 3 through CHL5 are taken from 
the benchmark sets by Hifi, which can be found at
$$
\mbox{\tt ftp://panoramix.univ-paris1.fr/pub/CERMSEM/hifi/2Dcutting/}.
$$
The sets {\em cgcut} and {\em gcut} are also guillotine-cut type
instances and can be found at the OR library.
Finally, we created five new OKP instances {\em okp} that are 
listed in detail in Table~\ref{okp1-5}.
They are random instances generated in the same
way as {\em beasley 1--12} after applying initial reduction.

A detailed listing of our results for these OKP-2 instances can be
found in Table~\ref{problems}. The first column lists the names of the
instances; the second shows the size of the container, followed by the
number of different box types and the total number of boxes. The fifth
column shows the number of nodes in the outer search tree, followed by
the total number of calls to the inner search tree, i.e., the times an
OPP had to be resolved.  The last two columns show the number of boxes
in an optimal solution, and the optimal value.  (Instance {\em gcut13}
is still unsolved; our lower bound corresponds to the best solution
found so far.)  Results are shown in Table~\ref{results}, where the
first column lists the instance names, the second column shows the
runtime on a PC with Pentium IV processor. Columns 3 and 4 give the
runtimes as reported by Beasley~\cite{BEA85} on a CYBER 855, by
Hadjiconstantinou and Christofides~\cite{HACHR95} on a CDC 7600.
In~\cite{CM04} Caprara and Monaci give runtimes for four different
algorithms. None of these algorithms dominates all the others; the best of them
(called $A_3$) uses a clever hybrid strategy for checking feasibility
during branching. 

The comparison in Table~\ref{results} should be considered
with some care, because different computers with different
compilers were used for the tests.
Some indication for the relative performance of the different machines
can be found at 
$$
\mbox{\tt http://www.netlib.org/benchmark/performance.ps,}
$$
where the results of the Linpack100 benchmark are presented
(see~\cite{DONG}). According to these results, a CDC 7600 
manages 120 Mflop/s, a CYBER 875 (Cyber 855 does not appear on the list)
gets 480 Mflop/s, a Sun Ultra SPARC achieves 7000 Mflop/s,
an Intel Pentium III (750 MHz) 13,800 Mflop/s, while
an Intel Pentium IV (2,8 GHz) manages 131,700 Mflop/s.
Note that these speeds may not be the same for other applications.
Furthermore, there is always a certain amount of chance involved when
comparing branch-and-bound procedures on individual instances.

Despite of these difficulties in comparison,
it is clear that our new method constitutes significant
progress. One indication is the fact that the ratio of running
times between ``large'' and ``small'' instances is smaller by several
orders of magnitude: As opposed to our two-level algorithm,
the search trees in the procedures by Beasley and 
Hadjiconstantinou/Christofides
appear to be reaching the threshold of exponential growth for 
some of the bigger instances. After 800 seconds, the procedure
by Hadjiconstantinou/Christofides timed out on instances
{\em beasley12}, {\em hadchr8}, and {\em hadchr11}, 
without finding a solution.
The comparison with Caprara and Monaci \cite{CM04} is less conclusive:
Both implementations fair pretty well on medium-sized instances,
with different behavior for large instances.
(Comparing a previous implementation of our algorithm with $A_3$,
Caprara and Monaci concluded that ``... the algorithm of [Fekete and Schepers]
appears to be more stable ...'') 
This behavior may also be the result of some differences
in branching strategies, which can always turn out differently
on individual instances. It should be noted that
the main basis for the success of our method is the underlying mathematical
characterization, and tuning of branching strategies
and bounds can be expected to provide further progress.
Promising may also be a combination of the
first-level strategy of \cite{CM04} with our second-level strategy.

At this stage, an instance like {\em cgcut13} is still out of reach,
even though we were able to improve the best known solution to
8,622,498, from 8,408,316 in \cite{CM04}, with an upper bound of 9,000,000,
leaving a gap of about 4\%.
It should be interesting to develop long-running,
special-purpose exact algorithms, just like Applegate et al.~\cite{13509}
did for the Traveling Salesman Problem.

\begin{table}[hp]
\begin{center}
\leavevmode
\caption{The new problem instances okp1--okp5.}
\label{okp1-5}
 
\begin{tabular}{|rl|}
\hline
Problem& okp1: container = (100,100), 15 box types (50 boxes)\\
$size$ =&[(4,90),(22,21),(22,80),(1,88),(6,40),(100,9),(46,14),(10,96),\\
&(70,27),(57,18),(10,84),(100,1),(2,41),(36,63),(51,24)]\\
$value$ =&[838,521,4735,181,706,2538,1349,1685,5336,1775,1131,129,179,\\
&6668,3551]\\
$n$ =&[5,2,3,5,5,5,3,1,3,1,1,5,5,2,4] \\
\hline
Problem& okp2: container = (100,100), 30 box types (30 boxes) \\
$size$ =&[(8,81),(5,76),(42,19),(6,80),(41,48),(6,86),(58,20),(99,3),(9,52),\\
&(100,14),(7,53),(24,54),(23,77),(42,32),(17,30),(11,90),(26,65),\\
&(11,84),(100,11),(29,81),(10,64),(25,48),(17,93),(77,31),(3,71),\\
&(89,9),(1,6),(12,99),(33,72),(21,26)]\\
value=& [953,389,1668,676,3580,1416,3166,537,1176,3434,676,1408,2362,\\
&4031,1152,2255,3570,1913,1552,4559,713,1279,3989,4850,299,\\
&1577,12,2116,2932,1214]\\
$n_{j}$ =& $1, \quad j \in \{1, \dots, 30\}$ \\
\hline
Problem& okp3: container = (100,100), 30 box types (30 boxes)\\
$size$ =&[(3,98),(34,36),(100,6),(49,26),(14,56),(100,3),(10,90),(23,95),\\
&(10,97),(50,47),(41,45),(13,12),(19,68),(50,46),(23,70),(28,82),\\
&(12,65),(9,86),(21,96),(19,64),(21,75),(45,26),(19,77),(5,84),\\
&(16,21),(23,69),(5,89),(22,63),(41,6),(76,30)]\\
value=&[756,2712,1633,2332,2187,470,1569,4947,2757,4274,4347,396,3866,\\
&5447,2904,6032,1799,929,5186,2120,1629,2059,2583,953,1000,\\
&2900,1102,2234,458,5458]\\
$n_{j}$ =& $1, \quad j \in \{1, \dots, 30\}$\\
\hline
Problem& okp4: container = (100,100), 33 box types (61 boxes)\\
$size$ =&[(48,48),(6,85),(100,14),(17,85),(69,20),(12,72),(5,48),(1,97),\\
&(66,36),(15,53),(29,80),(19,77),(97,7),(7,57),(63,37),(71,14),(3,76),\ \ \\
&(34,54),(5,91),(14,87),(62,28),(6,7),(20,71),(92,7),(10,77),(99,4),\\
&(14,44),(100,2),(56,40),(86,14),(22,93),(13,99),(7,76)]\\
$value$ =&[5145,874,2924,3182,2862,1224,531,249,6601,1005,6228,3362,907,\\
&473,6137,1556,313,4123,581,1999,5004,2040,3143,795,1460,841,\\
&1107,280,5898,2096,4411,3456,1406]\\
$n$ =&[1,2,1,1,1,1,3,3,2,1,3,1,1,2,2,1,3,1,2,1,3,3,1,1,2,3,2,3,2,1,1,3,3]\\
\hline
Problem& okp5: container = (100,100), 29 box types (97 boxes)\\
$size$ =&[(8,81),(5,76),(42,19),(6,80),(41,48),(6,86),(58,20),(99,3),(9,52),\\
&(100,14),(7,53),(24,54),(23,77),(42,32),(17,30),(11,90),(26,65),\\
&(11,84),(100,11),(29,81),(10,64),(25,48),(17,93),(77,31),(3,71),\\
&(89,9),(1,6),(12,99),(21,26)]\\
$value$ =&[953,389,1668,676,3580,1416,3166,537,1176,3434,676,1408,2362,\\
&4031,1152,2255,3570,1913,1552,4559,713,1279,3989,4850,299,\\
&1577,12,2116,1214]\\
$n$ =&[3,4,4,4,1,5,5,5,5,4,5,1,1,5,5,4,2,3,1,1,2,1,4,1,5,4,5,2,5] \\
\hline
\end{tabular}
\end{center}
\end{table}

\subsection{Generating New Test Instances for 2D and 3D}

\begin{table}[p]
\begin{center}
\begin{tabular}{|l|r|r|}
\hline
Class of& $w_{1}(x)$ evenly & $w_{2}(x)$ evenly \\
box types  & distributed on & distributed on \\ [0.5ex]
\hline
1 (Bulky in $2$) & $[1,50]$ & $[75,100]$ \\
2 (Bulky in $1$) & $[75,100]$ & $[1,50]$ \\
3 (Large) & $[50,100]$ & $[50,100]$ \\
4 (Small) & $[1,50]$ & $[1,50]$ \\
\hline
\end{tabular}

\vskip24pt

\begin{tabular}{|c|c|c|c|c|}
\hline
& \multicolumn{4}{|c|}{ Classes of box types } \\ \cline{2-5}
Instance type  & 1 & 2 & 3 & 4 \\ [0.5ex]
\hline  
I & 20 \%  & 20 \%  & 20 \%  & 40 \% \\
II & 15 \%  & 15 \%  & 15 \%  & 55 \% \\
III & 10 \%  & 10 \%  & 10 \%  & 70 \% \\
\hline
\end{tabular}
\end{center}
\caption{Random generation of OKP-$2$ test instances}
\label{genrand2d}
\end{table}

\begin{table}[p]
\begin{center}
\begin{tabular}{|l|r|r|r|}
\hline
Class of& $w_{1}(x)$ evenly & $w_{2}(x)$ evenly  & $w_{3}(x)$ evenly \\
box types  & distributed on & distributed on & distributed on \\ [0.5ex]
\hline
1 (bulky in $2$,$3$) & $[1,50]$ & $[75,100]$ &  $[75,100]$ \\
2 (bulky in $1$,$3$) & $[75,100]$ & $[1,50]$ & $[75,100]$ \\
3 (bulky in $1$,$2$) & $[75,100]$  & $[75,100]$ & $[1,50]$ \\
4 (large) & $[50,100]$ & $[50,100]$ & $[50,100]$  \\
5 (small) & $[1,50]$ & $[1,50]$  & $[1,50]$\\
\hline
\end{tabular}

\vskip24pt

\begin{tabular}{|c|c|c|c|c|c|}
\hline
& \multicolumn{5}{|c|}{ Class of box types } \\ \cline{2-6}
Instance type  & 1 & 2 & 3 & 4 & 5\\ [0.5ex]
\hline  
I & 20 \%  & 20 \%  & 20 \%  & 20 \%  & 20 \% \\
II & 15 \%  & 15 \%  & 15 \%  & 15 \% & 40 \% \\
III & 10 \%  & 10 \%  & 10 \%  & 10 \% & 60 \% \\
\hline
\end{tabular}
\end{center}
\caption{Random generation of OKP-$3$ test instances}
\label{genrand3d}
\end{table}

In order to get a broader test basis, and also include the three-dimensional case,
we generated 300 new test instances.
We followed the method described in \cite{MAVI96} and \cite{MAPV97}.

Our test instances are characterized by three parameters:
\begin{enumerate}
\item type of the instance (I, II, III) 
(see Tables~\ref{genrand2d} and \ref{genrand3d}),
\item number $m$ of box types, 
\item number $\nu$ of boxes for each box type. 
\end{enumerate}
Each of the instances consists of a container of size $100$ in each
coordinate direction, and $m$ box types, which are obtained as follows:

There are four (OKP-$2$) or five (OKP-$3$) classes of box types.
The type of the instance determines the probability of
each new box type $T_{t}$ to belong to one of these classes.
We use the distributions shown in 
Table~\ref{genrand2d} and Table~\ref{genrand3d}.                      

Depending on its class, the sizes of a box type are generated
randomly, according to the distributions in Table~\ref{genrand2d} 
and Table~\ref{genrand3d}. We round up to integer values.
In order to get the value of a box type,
the volume is multiplied with a random number from
$\{1,2,3\}$. The number of boxes in a new box type
it determined by the
parameter $\nu$, independent of $t$.

In this manner, we generated (for two as well as for three dimensions)
ten OKP instances for each of the three instance types and each of the
five parameter combinations:
$$
(m,\nu) \in \{ (20,1),(30,1),(40,1),(20,3),(20,4). \}
$$

\subsection{Results for New Test Instances}

\begin{table}[p]
\begin{center}
\begin{tabular}{|c|r|r|r|r|r|r|r|r|r|}
\hline &     &     & Solved & \multicolumn{3}{|c}{\# OKP nodes } & 
\multicolumn{3}{|c|}{ \# OPP nodes } 
\\ \cline{5-10}
Class & $m$ & $|V|$ & (out of 10) & Min & Av & Max & Min & Av & Max \\ 
\hline
& 20 & 20 &  10 & 5 & 57 & 174 & 0 & 8 & 23  \\
& 30 & 30 & 10 & 19 & 307 & 914 & 0 & 158 & 969 \\
I & 40 & 40  & 9 & 40 & 933 & 3826 & 0 & 431 & 2411 \\
& 20 & 60 &  9 & 31 & 231 & 677 & 1 & 287 & 1018   \\
& 20 & 80 &  8 & 82 & 336 & 943 & 234 & 147864 & 727719   \\
\hline
& 20 & 20 &  10 & 6 & 139 & 1038 & 0 & 305 & 2699 \\
& 30 & 30 &  10 & 28 & 548 & 1568 & 0 & 2873 & 26866  \\
II & 40 & 40 &  8 & 32 & 5062 & 28754 & 2 & 14910 & 84975  \\
& 20 & 60 &  7 & 36 & 297 & 571 & 5 & 237144 & 1633573   \\
& 20 & 80 &  6 & 62 & 536 & 1110 & 83 & 168530 & 280688   \\
\hline
& 20 & 20 & 10 & 3 & 117 & 516 & 0 & 299 & 1169  \\
& 30 & 30 &  10 & 82 & 737 & 1860 & 3 & 10588 & 53510  \\
III & 40 & 40 & 9 & 342 & 3865 & 10655 & 745 & 62065 & 416200 \\
& 20 & 60 &  8 & 31 & 1006 & 4064 & 3 & 345130 & 1174938   \\
& 20 & 80 &  2 & 96 & 196 & 296 & 241 & 85729 & 171218   \\
\hline
\end{tabular}
\end{center}
\caption{Results for randomly generated OKP-$2$ instances}
\label{okp2res}
\end{table}

\begin{table}[hp]
\begin{center}
\begin{tabular}{|c|r|r|r|r|r|r|}
\hline &     &    &  Solved & 
\multicolumn{3}{|c|}{ runtime/s } 
\\ \cline{5-7}
Class & $m$ & $|V|$ &  (out of 10) & Min & Av & Max \\ 
\hline
& 20 & 20 &  10 & 0.06 &  0.56 &  1.26 \\
& 30 & 30 &  10 & 0.29 &  4.48 & 13.59 \\
I & 40 & 40 &  9 & 1.31 & 22.02 & 76.30 \\
& 20 & 60 &  9 &  0.43 &  2.35 &  5.95  \\
& 20 & 80 &  8 &  0.99 & 62.46 & 243.41  \\
\hline
& 20 & 20 &  10  &  0.06 &  2.18 & 17.69 \\
& 30 & 30 &  10 &  0.36 & 10.64 & 39.59  \\
II & 40 & 40  & 8 &  0.55 & 51.12 & 152.46 \\
& 20 & 60 &  7 &  0.45 & 95.44 & 640.47  \\
& 20 & 80 &  6 &  1.86 & 112.89 & 267.90  \\
\hline
& 20 & 20 &  10 &  0.08 &  1.48 &  5.77  \\
& 30 & 30 &  10 &  1.07 & 17.67 & 53.00 \\
III & 40 & 40 & 9 &  6.66 & 103.10 & 313.91  \\
& 20 & 60 &  8 &  0.36 & 191.98 & 719.67  \\
& 20 & 80 &  2 &  2.18 & 34.52 & 66.86  \\
\hline
\end{tabular}
\end{center}
\caption{Runtimes for randomly generated OKP-$2$ instances using a Sun Ultra SPARC with 175MHz.}
\label{okp2zeit}
\end{table}

\begin{table}[hp]
\begin{center}
\begin{tabular}{|c|r|r|r|r|r|r|r|r|r|}
\hline
       &     &     & Solved & \multicolumn{3}{|c}{\# OKP nodes } & 
\multicolumn{3}{|c|}{ \# OPP nodes} 
\\ \cline{5-10}
Class & $m$ & $|V|$ & (out of 10) & Min & Av & Max & Min & Av & Max \\ 
\hline
& 20 & 20 & 10 & 1 & 73 & 352 & 0 & 22 & 82   \\
& 30 & 30 & 10 & 11 & 276 & 1190 & 1 & 59 & 291   \\
I & 40 & 40 & 10 & 73 & 953 & 2848 & 5 & 2684 & 20975   \\
& 20 & 60 & 10 & 20 & 541 & 2961 & 3 & 19896 & 198091   \\
& 20 & 80 & 9 & 42 & 414 & 1511 & 14 & 145 & 399   \\
\hline
& 20 & 20 & 10 & 11 & 75 & 328 & 1 & 35 & 166   \\
& 30 & 30 & 10 & 5 & 327 & 972 & 0 & 6579 & 62827   \\
II & 40 & 40 & 8 & 59 & 2197 & 13064 & 20 & 85465 & 671934   \\
& 20 & 60 & 5 & 1 & 292 & 719 & 0 & 232 & 912   \\
& 20 & 80 & 3 & 142 & 149 & 161 & 23 & 46 & 65   \\
\hline
& 20 & 20 & 10 & 5 & 57 & 138 & 0 & 4433 & 36747   \\
& 30 & 30 & 6 & 1 & 859 & 2250 & 1 & 3794 & 10063   \\
III & 40 & 40 & 3 & 17 & 652 & 1715 & 7 & 1326 & 3885   \\
& 20 & 60 & 3 & 51 & 3728 & 10842 & 27 & 55164 & 165276   \\
& 20 & 80 & 1 & 73 & 73 & 73 & 38 & 38 & 38   \\
\hline
\end{tabular}
\end{center}
\caption{Results for randomly generated OKP-$3$ instances.}
\label{okp3res}
\end{table}

\begin{table}[hp]
\begin{center}
\begin{tabular}{|c|r|r|r|r|r|r|}
\hline
 &  &  & Solved &
\multicolumn{3}{|c|}{ runtime/s } 
\\ \cline{5-7}
Class & $m$ & $|V|$ & (out of 10) & Min & Av & Max \\ 
\hline
& 20 & 20 & 10 &  0.06 &  1.63 &  7.76  \\
& 30 & 30 & 10 &  0.36 &  9.15 & 43.58  \\
I & 40 & 40 & 10 &  2.66 & 44.99 & 121.96  \\
& 20 & 60 & 10 &  0.50 & 18.33 & 125.76  \\
& 20 & 80 & 9 &  0.67 & 10.76 & 37.04  \\
\hline
& 20 & 20 & 10 &  0.26 &  1.76 &  6.92  \\
& 30 & 30 & 10 &  0.37 & 18.94 & 81.69  \\
II & 40 & 40 & 8 &  2.26 & 133.48 & 845.70  \\
& 20 & 60 & 5 &  0.28 & 12.00 & 38.94  \\
& 20 & 80 & 3 &  4.13 &  5.43 &  6.95  \\
\hline
& 20 & 20 & 10 &  0.29 &  4.73 & 21.69  \\
& 30 & 30 & 6 &  0.26 & 35.69 & 101.61  \\
III & 40 & 40 & 3 &  2.01 & 29.66 & 78.53  \\
& 20 & 60 & 3 &  1.21 & 211.63 & 607.83  \\
& 20 & 80 & 1 &  2.06 &  2.06 &  2.06  \\
\hline
\end{tabular}
\end{center}
\caption{Runtimes for randomly generated OKP-$3$ instances on a Sun Ultra SPARC with 175MHz, timeout after 1000s.}
\label{okp3zeit}
\end{table}

Tables~\ref{okp2res}, \ref{okp2zeit}, \ref{okp3res} and \ref{okp3zeit}
show the results for test runs on two- and three-dimensional
instances. For ten test instances of any combination of parameters,
we show how many of these instances we could solve within a
time limit of 1000 seconds on a Sun Ultra SPARC with 175 MHz. From the solved instances, 
we show the minimum (Min), the average (Av), and the maximum (Max)
of the number of OKP and OPP nodes, as well as the resulting runtimes.

It is evident that the difficulty grows with the percentage of
``small'' boxes. This is not very surprising, because these
boxes do not restrict the possibilities for the rest of a selected
subset as much as large or bulky boxes do.

The large difference in difficulty for instances with identical 
parameterization does not arise from our method of generation instances,
but is characteristic for instances of hard combinatorial optimization problems.
This effect has been known even for one-dimensional packing problems,
which have a much simpler structure. Because of this spread, the number
of nodes and runtimes are only significant for combinations of 
parameters where most instances could be solved.

For the OKP-$2$ with $m \le 40$ and $|V| \le 40$, we
could find an optimal solution in tolerable runtime for almost
all instances. For $60$ and more boxes, classes II and III
started to have higher numbers of instances that could not be solved
within the time limit. Only for instances with $80$ boxes
and about 70 \% of small boxes, our algorithm seemed
to reach its limits for the current implementation.

Even when taking into account that classes of three-dimensional
instances vary more with respect to the percentage of
small boxes than those for two dimensions, it is remarkable
that this percentage makes a huge difference with respect to the
difficulty of the resulting instances.
For an average of 20 \% of small boxes (Class I),
all instances (with the exception of a single one with $|V|=80$)
could be solved.
For an average of 40 \% of small boxes (Class II),
our method works well, at least for instances with $m \le 40, |V| \le 40$.
If the percentage of small boxes rises to 60 \% in Class III,
then even for $m = 30, |V| = 30$, our program does not
find an optimal solution for a large number of instances.

Summarizing, we can say that our new method has
greatly increased the size of instances that are practically solvable.
In particular, the size of the container is no longer a limiting factor.
It should be noted that even for three-dimensional instances 
with $m=20$, the 0-1 programs following the approach
by Beasley and Hadjiconstantinou/Christofides contain several 100,000
variables, even making the generous assumption of a grid reduction to
10 \%.

\subsection{A New Library of Benchmark Instances}
We are in the process of setting up a new library 
for multi-dimensional packing problems, called PackLib$^2$~\cite{packlib2}.
The idea is to have one place where benchmark instances, results, and solution
history can be found. For this purpose, we are using a universal XML-format
that allows inclusion of all this information. We provide parsers
for conversion directly into C formats, and converters for all
standard data formats. Results include visualization of solutions
by drawings of the feasible packings. Finally, we hope to provide
a number of algorithms at the website. Interested researchers are encouraged
to contact the first author.

\section{Solving Other Types of Packing Problems}
\label{other}
\subsection{Strip Packing Problems}
In an exact SPP procedure, we start with a heuristic for generating
a packing; its ``height'' is used as an upper bound.
A lower bound $\underline{h}$ can be obtained with the help of
the methods described in our paper \cite{pack2}. 
If there is a gap between these
bounds, we have to use enumeration.
Because the OPP is the decision version of the SPP
for a fixed objective value, an obvious approach would be binary search
in combination with the OPP algorithm from Section~\ref{oppsec}.

A more efficient method can be obtained by observing that
any OPP node that did not find a solution for height $h$
cannot possibly find a solution for height $h'<h$.
Thus, we can solve the SPP with the help of a modified version
of our OPP routine.

For finitely many boxes, there are only finitely many
possibilities for the minimal height of a packing.
The set $H$ of these values can be determined by using the
method from \cite{CHWH77} for computing normalized coordinates.

The height of the packing obtained by the heuristic is stored
under $h$. The variable $h'$ is initialized with the largest 
value from $H$ below $h$. 
We start the OPP tree search for the container with height
$h'$. If the algorithm finds a feasible packing, then $h$ is updated
to the value $h'$, and $h'$ is replaced by the next smaller value of $H$. 
Now the OPP search is done for container height $h'$. As noted above,
no search node that was dismissed before has to be considered again.
The search is performed until all search nodes have been checked,
or $h$ reaches the value $\underline{h}$ of the lower bound.

\subsection{Orthogonal Bin Packing Problems}
The basic scheme of our exact method 
follows the outline by 
Martello and Vigo~\cite{MAVI96}, and 
Martello, Pisinger, and Vigo~\cite{MAPV97}:

Within a branch-and-bound framework,
a packing (for a number of containers) is produced iteratively.
A list $L$ maintains all containers that are used.
In the beginning, $L$ is empty.
At each branching step, a box $b$ is either assigned to a container
$C$ in $L$, or a new container is generated for $b$, and added
to $L$. The crucial step is to check whether a container $C$ 
can hold all boxes that are assigned to it.

We get upper bounds by packing the unassigned boxes heuristically.
Our new suggestions concern the other steps of the approach,
which cause the largest computational effort:
\begin{enumerate}
\item computing lower bounds
\item solving the resulting OPPs.
\end{enumerate}
The improvement of the lower bounds from
\cite{MAVI96} and \cite{MAPV97} have already been discussed
in our paper~\cite{pack2}.

In \cite{MAVI96}, the resulting OPPs are enumerated by using the
method of Hadjiconstantinou/Christofides. For solving the three-dimensional
OPPs in \cite{MAPV97}, there is a special enumeration scheme,
using the principle of placement points described in Section~\ref{bbsec}.
As discussed in our paper~\cite{pack1b,pack1}, we get a drastic improvement
by using our method from Section~\ref{oppsec} that is based on 
packing classes.

\section{Conclusion}
In this paper, we have shown that higher-dimensional packing problems
of considerable size can be solved to optimality in reasonable
time, by making use of a structural characterization of feasible packings. 
Further progress may be achieved by refined lower bounds
and by using a more sophisticated outer tree search, as in 
the recent paper by Caprara and Monaci~\cite{CM04}.
Currently, we are working on such a more advanced implementation,
motivated by ongoing research on reconfigurable computing,
We expect this work to lead to progress for other problem variants.

\section*{Acknowledgments}
We thank 
Eleni Hadjiconstantinou who provided the data
for instances {\em hadchr8} and {\em hadchr12}.
Thank you to Alberto Caprara for some interesting discussions.
Two anonymous referees helped with a number of useful comments that improved
the overall presentation of this paper.



\end{document}